\documentclass[%
 reprint,
superscriptaddress,
 amsmath,amssymb,
 aps,
 prx,
]{revtex4-2}

\usepackage{graphicx}
\usepackage{dcolumn}
\usepackage{bm}
\usepackage[colorlinks]{hyperref}
\usepackage{physics}
\usepackage[squaren]{SIunits}
\usepackage{mathtools}
\usepackage{amsthm}
\usepackage{xcolor}
\usepackage{mathrsfs}
\usepackage{times}
\usepackage[normalem]{ulem}
\usepackage{algorithm2e}
\usepackage{comment}

\hypersetup{
	colorlinks=true,
	linkcolor=red,
	filecolor=blue,      
	urlcolor=blue,
	citecolor=blue
}
\newcommand{\rev}[1]{{\textcolor{black}{#1}}}

\newcommand{\NS}{S}
\newcommand{\NQ}{L}
\newcommand{\NI}{N}

\newcommand{\UI}{\bm{u}}
\newcommand{\EC}{C_T}
\newcommand{\ETC}{\bar{\mathcal{T}}}
\newcommand{\regmat}{\widetilde{\mathbf{F}}_N}
\newcommand{\RM}{\mathbf{F}_N}

\newcommand{\cu}{\mathbf{\Sigma}}  
\newcommand{\ci}{\mathbf{V}}       
\newcommand{\gr}{\mathbf{G}}       
\newcommand{\gh}{\mathbf{\Lambda}}  
\newcommand{\nsr}{\mathbf{R}}      

\begin{document}

\title[]{Quantifying the Expressive Capacity of Quantum Systems: Fundamental Limits and Eigentasks}



\makeatletter

\author{Fangjun Hu}
\thanks{These two authors contributed equally}
\affiliation{Department of Electrical and Computer Engineering, Princeton University, Princeton, NJ 08544, USA}

\author{Gerasimos Angelatos}
\thanks{These two authors contributed equally}
\affiliation{Department of Electrical and Computer Engineering, Princeton University, Princeton, NJ 08544, USA}
\affiliation{Raytheon BBN, Cambridge, MA 02138, USA}

\author{Saeed A. Khan}
\affiliation{Department of Electrical and Computer Engineering, Princeton University, Princeton, NJ 08544, USA}

\author{Marti Vives}
\affiliation{Department of Electrical and Computer Engineering, Princeton University, Princeton, NJ 08544, USA}
\affiliation{Q-CTRL, Santa Monica, CA 90401, USA}

\author{Esin T\"ureci}
\affiliation{Department of Computer Science, Princeton University, Princeton, NJ 08544, USA}

\author{Leon Bello}
\affiliation{Department of Electrical and Computer Engineering, Princeton University, Princeton, NJ 08544, USA}

\author{Graham E. Rowlands}
\affiliation{Raytheon BBN, Cambridge, MA 02138, USA}

\author{Guilhem J. Ribeill}
\affiliation{Raytheon BBN, Cambridge, MA 02138, USA}

\author{Hakan E. T\"ureci}
\affiliation{Department of Electrical and Computer Engineering, Princeton University, Princeton, NJ 08544, USA}

\date{\today}

\begin{abstract}
The expressive capacity of quantum systems for machine learning is limited by quantum sampling noise incurred during measurement. Although it is generally believed that noise limits the resolvable capacity of quantum systems, the precise impact of noise on learning is not yet fully understood. We present a mathematical framework for evaluating the available expressive capacity of  general quantum systems from a finite number of measurements, and provide a methodology for extracting the extrema of this capacity, its eigentasks. Eigentasks are a native set of functions that a given quantum system can approximate with minimal error. We show that extracting low-noise eigentasks leads to improved performance for machine learning tasks such as classification, displaying robustness to overfitting. We obtain a tight bound on the expressive capacity, and present analyses suggesting that correlations in the measured quantum system enhance learning capacity by reducing noise in eigentasks. These results are supported by experiments on superconducting quantum processors. Our findings have broad implications for quantum machine learning and sensing applications.
\end{abstract}

\maketitle



\section{Introduction}

\begin{figure}
    \centering
    \includegraphics[width=\columnwidth]{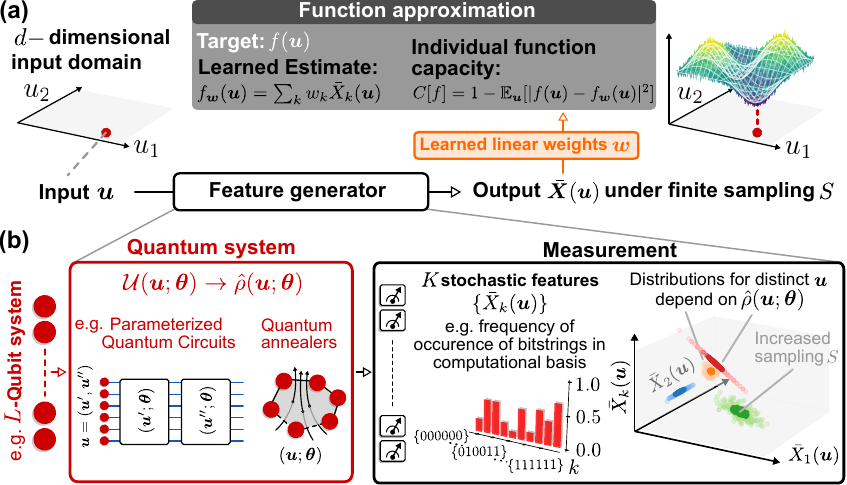}
    \caption{(a) Representation of the learning framework considered in this work: inputs $\UI$ are transformed to a set of outputs via a parameterized feature generator, here implemented using a finitely-sampled quantum system as shown in (b). Inputs are encoded in the state of a quantum system via a general quantum channel $\mathcal{U}$,
    \rev{and information is then extracted through a positive operator-valued measure}. This framework describes a wide range of practical quantum systems, from quantum circuits used in supervised \rev{or generative} quantum machine learning, to quantum annealers exhibiting continuous evolution, and beyond, all defined by a general quantum channel with parameters $\bm{\theta}$. Extracted information takes the form of $K$ stochastic features $\bar{\bm{X}}$ obtained under finite shots $\NS$. The geometric structure of distributions of these measured features is fundamentally determined by quantum sampling noise, which depends on the quantum state $\hat{\rho}(\UI;\bm{\theta})$, and hence on the nature of the mapping from input $\UI$ to this quantum state. We show four obtained distributions differing only in the values of inputs $\UI$ to highlight this dependence. As shown in (a), learned estimates for desired functions are then constructed via a linear combination ${\bm{w}}$ of $\bar{\bm{X}}$, with a resolution limited by $\NS$. Capacity $C[f]$ then quantifies the error in the approximation of a target function $f$ via this scheme. 
    }
    \label{fig:NISQRC_Schematic}
\end{figure}

Learning with quantum systems is a promising application of near-term quantum processors, with several recent demonstrations in both quantum machine learning (QML)~\cite{grant_hierarchical_2018, havlicek_supervised_2019, tacchino_quantum_2020, chen_temporal_2020, suzuki_natural_2022, rudolph_generation_2022} and quantum sensing~\cite{meyer_fisher_2021, ma_adaptive_2021, Marciniak_Monz_2022}. A broad class of such data-driven applications proceed by embedding data into the evolution of a quantum system, where the embedding, dynamics, and extracted outputs via measurement are all governed by a set of general parameters $\bm{\theta}$ \cite{benedetti_parameterized_2019,cerezo_variational_2021, schuld_machine_2021}. 
Depending on the learning scheme, different components $\bm{\theta}$ of this general framework may be trained for optimal performance of a given task. In all cases the fundamental role of the quantum system is that of a high-dimensional feature generator: given inputs $\bm{u}$, a set of frequencies for the occurrence of different measurement outcomes act as a parameterized feature vector implementing a function $f (\UI)$ that minimizes a chosen loss function (see Fig.\,\ref{fig:NISQRC_Schematic}). The relationship between the physical structure of the model and the function classes that can be expressed with high accuracy is a fundamental question of basic importance to the success of quantum models. Recent results have begun to shed light on this important question and provide guidance on the choice of parameterized quantum models~\cite{wright_capacity_2019, Du2020, du_efficient_2022, sim_expressibility_2019, wu_expressivity_2021, larose_robust_2020, schuld_effect_2021, Abbas_power_2021,  Holmes2022}.
Yet when it comes to experimental implementations, the presence of noise is found to substantially curtail theoretical expectations for performance~\cite{grant_hierarchical_2018, havlicek_supervised_2019, tacchino_quantum_2020}.

Given an input $\bm{u}$ to a general dynamical system, we define its Expressive Capacity (EC) as a measure of the accuracy with which $K$ linearly independent functions $\{f(\UI)\}$ of the input can be constructed from $K$ measured features. This is a suitable generalization  of the Information Processing Capacity introduced in Ref.~\cite{dambre_information_2012} to noisy systems, and a direct quantification of the information content of these measured features. A central challenge in determining the EC for \textit{quantum} systems is the fundamentally stochastic nature of measurement outcomes. Even when technical noise due to system parameter fluctuations is minimized as in an error-corrected quantum computer, there is a fundamental level of noise, the quantum sampling noise (QSN), which cannot be eliminated in learning with quantum systems. Quantum sampling noise therefore sets a fundamental limit to the EC of any physical system. Although QSN is well-understood theoretically, a formulation of its impact on learning is a challenging task as it is strongly determined by the quantum state of the system relative to the measurement basis, and is highly correlated when quantum coupling is present. Consequently, the impact of QSN is often ignored~\cite{fujii_harnessing_2017, benedetti_parameterized_2019, cerezo_variational_2021,schuld_machine_2021, martinez-pena_information_2020} (with a few exceptions~\cite{wright_capacity_2019, arrasmith_effect_2021, garcia-beni_scalable_2022, anshu_sample-efficient_2021}), even though it can place strong constraints on practical optimization~\cite{arrasmith_effect_2021} and performance~\cite{garcia-beni_scalable_2022}. 

In this article, we develop a mathematical framework to quantify the EC that exactly accounts for the structure of QSN, providing a tight bound for a quantum system with $K$ measurement outcomes under $S$ samples,
and illustrate how a mathematical framework for its quantification can guide experimental design for QML applications. While the strength of the EC lies in its generality, we provide numerical examples and experimental results confirming that higher EC is typically indicative of improved performance on specific tasks. As such, the EC provides a metric to guide ans\"atze-design for improved learning performance in a task-agnostic and parameter-independent manner. Specifically, our work identifies enhancement in measurable quantum correlations as a general principle to increase the EC of quantum systems under finite sampling.  

Our work goes beyond simply defining the EC as a figure of merit for parameterized quantum systems, however. In particular, we offer a reliable methodology to identify the native function set that is most accurately resolvable by a given encoding under finite sampling parameterization under QSN.
Equivalently, we show that this defines a construction of measured features spanning the accessible information which is optimally robust to noise in readout, thereby furnishing a critical tool employable in any QML framework to improve learning in experimental settings.
This strategy for defining the noise-constrained EC naturally focuses on accessible noisy output features under a specified measurement scheme, as opposed to unmeasured degrees of freedom. This makes the EC an efficiently-computable quantity in practice, as we demonstrate using both numerical simulations and experiments on IBM Quantum's superconducting multi-qubit processors~\cite{IBMQ_2022}.



\rev{\section{Theoretical Analysis}} 

\rev{\subsection{Quantum Sampling Noise and Learning}}

The most general approach to learning from data using a generic quantum system is depicted schematically in Fig.\,\ref{fig:NISQRC_Schematic}. A table with symbols and abbreviations used in the text can be found in Appendix \ref{app:table}. 
Any quantum learning scheme begins with embedding the data $\bm{u}$ through a quantum channel parameterized by $\bm{\theta}$ acting on a known initial state, 
\begin{align}
\hat{\rho}(\UI;\bm{\theta}) = \mathcal{U}(\UI;\bm{\theta} )\hat{\rho}_0.
\label{eq:qchannel}
\end{align}
\rev{This channel includes all quantum operations applied to the input data; to obtain the computational output or perform further classical processing, one must extract information from the quantum system via a set of measurements described most generally as a positive operator-valued measure (POVM). Specifically, we define a set of $K$ POVM elements \{$\hat{M}_k$\}, each associated with a distinct measurement outcome indexed $k$, and 
constrained only by the normalization condition $\sum_{k=0}^{K-1}\hat{M}_k = \mathbf{I}$ (and hence not necessarily commuting).} 

A single measurement or ``shot'' \rev{ then yields a discrete index $k^{(s)}(\UI)$ specifying the observed outcome: for input $\UI$, if outcome $k$ is observed in shot $s$ then $k^{(s)}(\UI) \gets k$}. Measured features are then constructed by ensemble-averaging over $\NS$ repeated shots:
\begin{align}
    \bar{X}_k(\UI) = \frac{1}{\NS} \sum_s \delta(k^{(s)}(\UI), k)
    \label{eq:Xsum}
\end{align}
Hence $\bar{X}_k(\UI)$ in this case is the empirical frequency of occurrence of the \rev{outcome $k$} in $S$ repetitions of the experiment with the same input $\UI$. These measured features are formally random variables that are unbiased estimators of the expected value of the corresponding element $\hat{M}_k$ as computed from $\hat{\rho}({\UI})$
. Explicitly 
\begin{align}
    {\rm lim}_{S\to\infty} \bar{X}_k({\UI}) = x_k({\UI}) \equiv {\rm Tr}\{\hat{M}_k \hat{\rho}({\UI};\bm{\theta})\},
    \label{eq:mapping}
\end{align}
so that $x_k$ is the probability of occurrence of the $k$th \rev{outcome as specified by the quantum state.  These probability amplitudes encompass the accessible information in $\hat{\rho}({\UI};\bm{\theta})$: any observable under this set can be written as a linear combination of POVM elements $\hat{O}_{\bm{W}} = \sum_k W_k \hat{M}_k$, such that $\langle \hat{O}_{\bm{W}} \rangle = \bm{W}^T \bm{x}$.}

In QML theory, it is standard to consider the limit $\NS \to \infty$, and to thus use expected features $\{x_k(\UI)\}$ for learning. In any actual implementation however, measured features $\{\bar{X}_k(\UI)\}$ must be constructed under finite $\NS$, in which case their fundamentally quantum-stochastic nature can no longer be ignored. More precisely, $\bar{\bm{X}}$ are samples from a multinomial distribution with $S$ trials and $K$ categories, which can be decomposed into their expected value -- the quantum-mechanical event probabilities $\bm{x}$ -- and a zero-mean, input-dependent noise term $\boldsymbol{\zeta}(\UI)$:
\begin{equation}
    \bar{\bm X}(\UI) =  {\bm x}(\UI) +\frac{1}{\sqrt{S}}  \boldsymbol{\zeta}(\UI),
    \label{eq:xbar}
\end{equation}
\rev{Here $\boldsymbol{\zeta}$ encodes the multinomial statistics of QSN; it has non-zero cumulants of all orders, of which the covariances take the particular $S$-independent form},
\begin{align}
    {\rm Cov}[{\zeta}_j, {\zeta}_k](\UI) 
    \equiv
    \cu_{jk}(\UI)
    = \delta_{jk} x_k(\UI) - x_j(\UI) x_{k}(\UI)
    \label{eq:cov}
\end{align}
or more concisely $\cu = \mathrm{diag}(\bm{x})-\bm{x}\bm{x}^T$. \rev{We note that the above expressions are exact; the factor of $1/\sqrt{\NS}$ is merely extracted for convenience of the analysis to follow, and in particular is \textit{not} meant to suggest an expansion for large $\NS$; cumulants of $\boldsymbol{\zeta}$ beyond second-order inherit a complicated $S$-dependence} \cite{Wishart1949}. 


\rev{Before developing our capacity analysis, we note that when viewed in isolation Eq.\,\eqref{eq:xbar} defines an extremely general map between inputs $\UI$ and outputs $\bar{\bm X}(\UI)$ assembled from $S$ measurements. The readout features it describes could therefore have been extracted from \textit{any} dynamical system with stochastic outputs,
including systems that are entirely classical. 
This is no restriction -- Eq.\,(\ref{eq:xbar}) also applies to a very broad class of quantum systems: ultimately, measurement outcomes from quantum systems are also recorded by an observer as classical stochastic variables. Where, then, is the quantum nature of measured features apparent? This is encoded in the fact that for quantum systems all statistical properties of stochastic readout features $\bar{\bm X}(\UI)$ -- namely first-order cumulants $\bm{x}(\UI)$, second-order cumulants $\bm{\Sigma}(\UI)$, and all higher-order cumulants -- are determined explicitly by the quantum state $\hat{\rho}({\UI})$, which itself may be a distribution that is hard to generate classically. 
}


\rev{The framework we develop here allows characterization of the function-learning capacity of any noisy dynamical system satisfying Eq.\,\eqref{eq:xbar}, and provides a practical, experimentally applicable methodology to optimize learning by avoiding overfitting to noise in ML tasks. When applied to classical systems, it can be viewed as a means of statistical inference based on data containing classical noise~\cite{cox_principles_2006}. However, our primary interest is the study of quantum systems, where the fundamental model for the noise process depends nontrivially on the quantum state, a dependence we account for exactly. This enables us to extract the limits on function-learning capacity set by QSN and its dependence on the encoding. In this paper, using both theoretical studies and experiments on real quantum devices, we analyze how this capacity depends on quantum properties such as the degree of measured correlations, and how our framework can be applied for optimal learning in practical QML tasks such as classification. }

\rev{\subsection{Expressive Capacity}}

\rev{Returning to the situation depicted in Fig.\,\ref{fig:NISQRC_Schematic}, QML and quantum sensing can generically be cast as encoding data in a parameterized quantum system, and then using measurement outcomes to approximate a desired function $f(\UI)$ (here assumed to be square-integrable ${\mathbb E}_{\UI} [|f(\UI)|^2] < \infty$).}
The input data is defined with respect to a distribution $p(\UI)$ which can be continuous or discrete: ${\mathbb E}_{\UI} [f] \equiv \int \dd \UI \, p(\UI) f(\UI) \simeq \frac{1}{N}\sum_n f(\UI^{(n)})$ for i.i.d.~sampling obeying $\UI^{(n)} \sim p(\UI)$ for all $n \in [N]$. 
Recalling that features $\bar{\bm{X}}(\UI)$ are estimators of POVM expectation values -- the linear combination of which can be used to construct all accessible observables  -- $f(\UI)$ is approximated for finite $S$ as $f_{\bm{W}}(\UI) = \bm{W}^T \bar{\bm{X}}(\UI)$.
To quantify the fidelity of this approximation, we introduce the functional capacity~\cite{dambre_information_2012, wright_capacity_2019, martinez-pena_information_2020}, which is simply the normalized mean-squared accuracy 
of the estimate $f_{\bm{W}}$
\begin{align}
    C[f] =  1- \min_{\boldsymbol{W} \in \mathbb{R}^K} \frac{\mathbb{E}_{\UI}[|f(\UI) - f_{\bm{W}}(\UI)|^2]}{{\mathbb E}_{\UI} [|f(\UI)|^2]}. 
    \label{eq:fcap}
\end{align}
Minimizing error in the approximation of $f(\UI)$ by $f_{\bm{W}}(\UI)$ over the input domain to determine capacity thus requires finding ${\boldsymbol{w}} = \mathrm{argmin}_{\boldsymbol{W}} \mathbb{E}_{\UI} [|f - \bm{W}^T \bar{\bm{X}}|^2]$, which can be always be expressed analytically via a pseudoinverse operation (see Appendix \ref{sec:Information_capacity_saturation}). This function capacity is constructed such that $0 \leq C[f]\leq 1$.

The choice of a linear estimator and a mean squared error loss function may appear restrictive at first glance, but the generality of our formalism averts such limitations. The use of a linear estimator applied directly to readout features appears to preclude classical nonlinear post-processing of measurements; however, this is simply to ensure the calculated functional capacity is a measure of the parameterized quantum system itself, and not of
a classical nonlinear layer. Furthermore, the mean squared loss effectively describes the first term in a Taylor expansion of a wide range of arbitrary nonlinear post-processing and non-quadratic loss functions~(see Appendix \ref{app:ComplxNonLin}). 


To extend the notion of capacity to a task-independent 
metric representing how much classical information about an input can be extracted from a system in the presence of noise, we sum the function capacity over a basis of functions $\{ f_\ell \}_{\ell \in \mathbb{N}}$ which are complete and orthonormal with respect to the input distribution, i.e.\,equipped with the inner product $\langle f_{\ell}, f_{\ell'} \rangle_{p} = \int f_{\ell}(\UI) f_{\ell'}(\UI) p(\UI) \dd \UI = \delta_{\ell \ell'}$.  The \textit{total Expressive Capacity} (EC) is then
$
    \EC \equiv \sum_{\ell=0}^{\infty} C[ f_{\ell} ], 
$
which effectively quantifies how many linearly-independent functions can be expressed from a linear combination of $\{\bar{X}_k (\UI) \}$. Our main result -- proven in detail in Appendix \ref{app:EC} -- is that 
\rev{given any $S\in \mathbb{N}^{+}$}, the EC for a physical system whose measured features are stochastic variables of the form of Eq.\,(\ref{eq:xbar}) is given by
\begin{align}
    \!\!\EC(\bm{\theta}) 
    = \mathrm{Tr} \left(\! \left( \gr + \frac{1}{\NS} \ci \right)^{\!\! - 1}\!\! \gr \right) = \sum_{k=0}^{K-1} \frac{1}{1 + \beta_k^2(\bm{\theta})/\NS}. 
    \label{eq:EC}
\end{align}
The first equality, arrived at through straight-forward algebraic manipulation, is written in terms of the expected feature Gram and covariance matrices $\gr \equiv \mathbb{E}_{\UI} [\bm{x}\bm{x}^T]$ and $\ci \equiv \mathbb{E}_{\UI} [\boldsymbol\Sigma]$ respectively.  We later demonstrate that these expected quantities can be accurately estimated in experiment and consequently under finite $\NS$ (see Appendix \ref{app:tilde_correction} and Eq.\,(\ref{eq:btilde})). 
The second equality remarkably provides a closed-form expression for $\EC$ at any $S$, which is independent of the generally-infinite set $\{ f_\ell \}_{\ell \in \mathbb{N}}$ (and thus not subject to numerical challenges associated with its evaluation over such a set~\cite{dambre_information_2012}).
Instead the EC is entirely captured by the function capacity of $K$ distinct functions, and for a given physical system is fully characterized by the spectrum of eigenvalues $\{ \beta^{2}_k \}_{k \in [K]}$ satisfying the generalized eigenvalue problem
\begin{align}
    \ci \bm{r}^{(k)} = \beta_k^2 \gr \bm{r}^{(k)}. \label{eq:eigenprob}
\end{align}
In the above, all quantities depend on $\boldsymbol\theta$ and thus the specific physical system and input embedding via the Gram ($\gr$) and covariance ($\ci$) matrices.
Associated with each $\beta_k^2$ is an eigenvector $\bm{r}^{(k)}$ living in the space of measured features and thus defining a set of $K$ orthogonal functions via the linear transformation
\begin{align}
    y^{(k)} (\UI) = \sum_{j} r_{j}^{(k)} x_{j} (\UI)
\end{align}
We refer to $\{y^{(k)}\}$ as {\it eigentasks}, as they form the minimal set of orthonormal functions ($\mathbb{E}_{\UI} [y^{(j)} y^{(k)} ]  = \delta_{jk}$) which completely accounts for the EC of a physical system and thus the accessible information content present in its measured features. \rev{Specifically, the capacity to approximate a given $y^{(k)}$ with $S$ shots is $C[y^{(k)}]=1/(1+\beta^2_k/S)$: the EC in Eq.\,\eqref{eq:EC} is simply a sum of eigentask capacities.} This further highlights that a given parameterized system can only approximate a target function to the degree that it can be written as a linear combination of $\{y^{(k)}\}$.  The eigentasks thus serve as a powerful basis for learning, as shall be explored in Sec.\,\ref{sec:eigentasklearning}.

\rev{
Defining measured eigentasks $\bar{y}^{(k)}(\UI) = \sum_j {r}_j^{(k)} {\bar{X}}_j(\UI)$, we find (see Appendix \ref{sec:noisy_ET}) that $\{\bm{r}^{(k)}\}$ specify a unique linear transformation that simultaneously orthogonalizes not only the signal, but also the associated noise: $\mathbb{E}_{\UI} [\bar{y}^{(j)} \bar{y}^{(k)} ] = \delta_{jk}(1+\beta^2_{k}/S)$.} 
The term $\beta^2_k/S$ is thus the mean squared error, or noise power, associated with the approximation of eigentask $y^{(k)}$; equivalently,  $\bar{y}^{(k)}$ has a signal-to-noise ratio of $S/\beta^2_k$. This leads to a natural interpretation of $\{\beta_k^2\}$ as noise-to-signal (NSR) eigenvalues. The eigentasks, 
ordered in increasing noise strength $0 \leq \beta^2_0 \leq \beta^2_1 \leq \cdots \leq \beta^2_{K-1} < \infty$, are the orthogonal set of functions maximally robust to noise.

\rev{Having developed our framework for EC in the most general context, in the remainder of this paper we will use it to analyze \textit{quantum} systems in particular.} The same quantitative metrics -- EC, eigentasks, and NSR eigenvalues -- now carry the significance of being determined by an arbitrary parameterized quantum state $\hat{\rho}(\UI; \bm{\theta})$, whose data-dependence ideally can be hard to model classically. Our formulation of EC hence encompasses general quantum states, to the best of our knowledge the first of its kind, going beyond characterizations of noise-constrained capacity that have been attempted for linear classical systems~\cite{hermans_memory_2010} and Gaussian quantum systems~\cite{garcia-beni_scalable_2022}. The eigentasks then reveal the set of orthogonal functions best approximated by the quantum system, and hence are sensitive to properties such as the degree of quantum correlations. Finally, the fidelity of approximation of these native functions -- determined by NSR eigenvalues -- is constrained fundamentally by QSN.

From Eq.\,\eqref{eq:EC}  we have $\lim_{S\to\infty} \EC = {\rm Rank}\{\gr\}$, where ${\rm Rank}\{\gr\} = K$, the number of measured features, provided no special symmetries exist (see Appendix \ref{app:Function-independence}). This important result reveals that in the absence of noise all dynamical systems -- independent of pararameterization -- have a capacity which is simply the number of independent accessible degrees of freedom \cite{dambre_information_2012, hermans_memory_2010}
The generic exponential scaling of measured degrees of freedom with quantum system-size (e.g.\,$K=2^L$ for $L$-qubit systems subject to a computational basis measurement) is often-cited as a motivator for performing ML with quantum systems~\cite{Kalfus_2022, wright_capacity_2019, martinez-pena_information_2020}. 
\rev{However, as will be demonstrated shortly, the EC of quantum systems can be significantly reduced from this limit for finite $S$ in a way that strongly depends on the encoding. By evaluating the ability of quantum systems to accurately express functions in the presence of QSN, the capacity analysis above provides an important metric to asses the utility of quantum platforms for learning in practice.
 } 

\rev{\subsection{Expressive Capacity of Quantum 2-designs}}


\rev{
We first consider the EC of quantum 2-designs: systems with fixed $\bm{\theta}$ that map inputs to a unitrary ensemble $\{p(\UI) \dd \UI, \hat{U}(\UI;\bm{\theta})\}$ whose first and second moments agree with those from a uniform (Haar) distribution of unitaries. Quantum 2-designs are important to recent QML studies \cite{cerezo_variational_2021, Holmes2022} due to their role in defining ``expressibility'' \cite{sim_expressibility_2019, wu_expressivity_2021}: a metric quantifying how close a parameterized quantum system is to such a 2-design.
The capacity eigenproblem Eq.\,(\ref{eq:eigenprob}) for any quantum 2-design over $K$-dimensions can be solved analytically (see Appendix \ref{app:2designsol}), yielding a flat spectrum of NSR eigenvalues $\beta^2_k = K (1-\delta_{k0})$.  This results in an EC
\begin{align}
    C_T = K \cdot \frac{S + 1}{S + K},
\end{align}
which at finite $\NS$ can be significantly lower than $K$.  For qubit-based systems with $K=2^L$, all $k\neq 0$ eigentasks have a noise strength $2^L/\NS$, requiring $S$ to grow exponentially with qubit-number $L$ in order to extract useful features.
}

\rev{
A quantum 2-design is thought of as having maximal ``expressibility'', however we see that its EC always vanishes exponentially with system size for a fixed finite $S$.  It is exactly such systems that have been shown to lead to barren plateaus which preclude learning \cite{Holmes2022}. To emphasize the distinction with ``expressibility'', we note that EC reflects how much classical information can be extracted from  the entire ``quantum computational stack'' in practice: from an abstract algorithm, to the quantum hardware on which its implemented, and the classical electronics used for control and readout.  EC requires only noisy computational outputs $\{\bar{X}_k (\UI) \}$ and is thus efficiently-computable in experiment -- unlike more abstract metrics \cite{sim_expressibility_2019, wu_expressivity_2021, meyer_fisher_2021} -- yielding a directly relevant metric for learning with quantum hardware.
}

\section{Experimental Results} 
\label{sec:ibmq}

To demonstrate the practical utility of our framework, we now show how the spectrum $\{\beta_k^2\}$, the EC, and eigentasks can all be computed for real quantum devices in the presence of parameter fluctuations and device noise. 
{We {reiterate} at the outset that our approach for quantifying the EC of a quantum system is very general, and can be applied to a variety of quantum system models.} 
For practical reasons, we perform experiments on $L$-qubit IBM Quantum (IBMQ) processors, whose dynamics is described by a parameterized quantum circuit containing single and two-qubit gates. However, as an example of the broad applicability of our approach, in Appendix \ref{app:H-ansztz} we compute the EC for $L$-qubit quantum annealers via numerical simulations, governed by the markedly different model of continuous-time Hamiltonian dynamics.

\begin{figure}[t]
    \centering
    \includegraphics[width=\columnwidth]{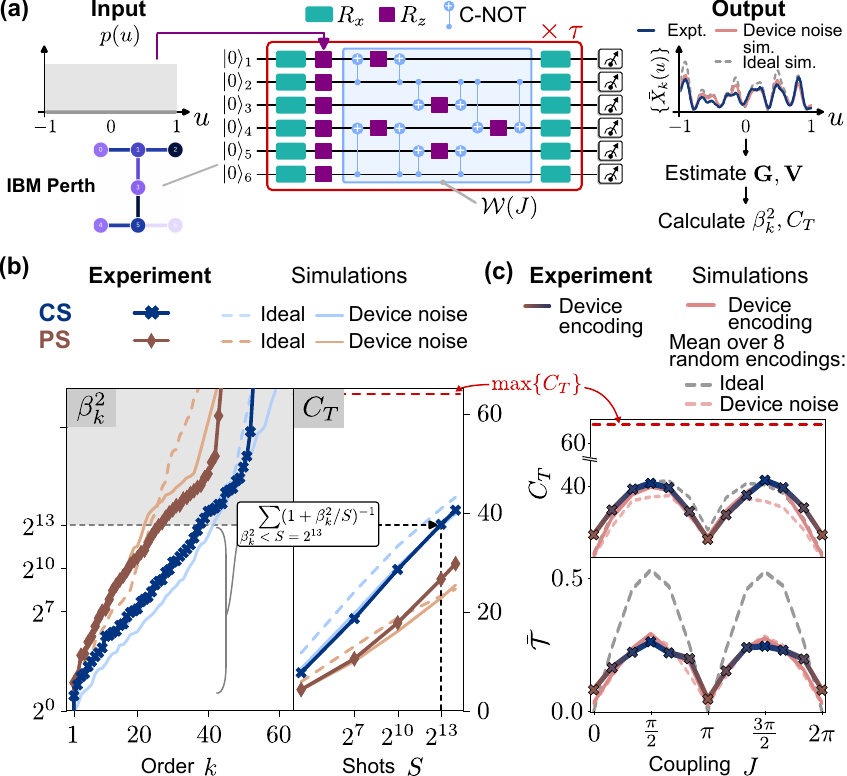}
    \caption{
    (a) A representation of the EC analysis, featuring the IBMQ Perth device and a schematic of the quantum circuit considered in this section.
    On the right, the specific feature plotted is $\bar{X}_1 (u)$ ($\bm{b}_1=000001$) with $S=2^{14}$ shots. 
    (b) Left panel: Device noise-to-signal spectrum $\beta^2_k$ for 
    a specific encoding as \rev{a correlated system (CS)}, $J=\pi/2$ (blue crosses) and product system (PS), $J=0$ (brown diamonds). Ideal (solid) and device noise (dashed) simulations are also shown.  Note the agreement between device and simulation, along with distortion from more direct exponential growth in $\beta^2_k$ with $k$ in the ideal case, due to device errors. Right panel: $\EC$ vs. $\NS$ calculated from the left panel. At a given $\NS$, the $\EC$ can be approximated by performing the indicated sum over all $\beta_k^2 < \NS$. (c) Expressive capacity $\EC$ (top panel) and expected total correlation $\bar{\mathcal{T}}$ (lower panel) for the chosen encoding under $S=2^{14}$ from the IBM device, and device noise simulations (dashed peach). Average metrics over 8 random encodings for device noise (solid peach) and ideal (solid gray) simulations are also shown. The $S\to\infty$ expressive capacity of these encodings always attains the ${\rm max}\{\EC\}=64$, indicated in dashed red. 
    }
    \label{fig:Genc1}
\end{figure}

On IBMQ devices, resource limitations restrict our computation of EC to 1D inputs $u$ that are uniformly distributed, $p(u)=\mathrm{Unif}[-1,1]$, see Fig.\,\ref{fig:Genc1}(a).
\rev{Specifically, we are limited to $N=300$ distinct inputs; a 1D distribution then ensures features $\{\bar{X}_k (\UI) \}$ are sufficiently densely sampled to approach the continuum limit, and are also easy to visualize.}
We emphasize that this analysis can be straightforwardly extended to multi-dimensional and arbitrarily-distributed inputs given suitable hardware resources, without modifying the form of the Gram and covariance matrices.

We are only now required to specify the model of the quantum system, and choose an ansatz  tailored to be natively implementable on IBMQ processors (see Appendix \ref{sec:NISQRC_Architecture_Detail}). 
\rev{We fix $\hat{\rho}_0 = \ketbra{0}{0}^{\otimes L}$; note, however, that any other initial state may be implemented via an additional unitary and absorbed into the ``encoding'', i.e.\,the quantum channel $\mathcal{U}(u;\bm{\theta})$ of Eq.\,\eqref{eq:qchannel}. In this way, the dependence of EC on initial states  could be explored in future studies.} 

The circuit we choose consists of $\tau \in \mathbb{N}$ repetitions of the same input-dependent circuit block depicted in Fig.\,\ref{fig:Genc1}(a). The block itself is of the form $\mathcal{R}_{x}(\bm{\theta}^x/2) \mathcal{W}(J)  \mathcal{R}_{z}(\bm{\theta}^z +\bm{\theta}^I u ) \mathcal{R}_{x}(\bm{\theta}^x/2) $, where $ \mathcal{R}_{x/z}$ are Pauli-rotations applied qubit-wise, e.g.\,$ \mathcal{R}_{z} = \prod_{l} R_{z}({\theta}^z_l +{\theta}^I _l u) $. A two-qubit coupling gate acts between physically connected qubits in the device and can be written as $\mathcal{W}(J) = \prod_{\langle l, l' \rangle} \mathrm{exp}\{- i \frac{J}{2} \hat{\sigma}^z_{l} \hat{\sigma}^z_{l'}\} $.
Within the structure of this ansatz, we will choose all single-qubit rotation parameters randomly: $\theta^{x/z}_l \sim \mathrm{Unif}[0,2\pi]$ and  $\theta^{I}_l\sim \mathrm{Unif}[0,10\pi]$, generally representing a circuit trained for a particular unspecified task. Each instance of random parameters, along with associated dissipative processes, specifies the quantum channel  $\mathcal{U}(u;\bm{\theta})$ which we refer to as an ``encoding''.  We will study the performance of an overall ansatz by looking at the behavior averaged across encodings as hyperparameters such as $J$ are varied.  In this work we also choose $\tau=3$, which limits circuit depth and associated prevalence of gate errors, while still generating a complex state with correlation generally distributed throughout all qubits.


Finally, we consider feature extraction via a computational basis measurement as is standard in quantum information processing: the POVM elements are the $K=2^L$ projectors $\hat{M}_k = \ketbra{\boldsymbol{b}_k }$, where $\boldsymbol{b}_k$ is the $L$-bit binary representation of the integer $k$. However, as with state preparation, measurements in any other basis can be (and in practice, are) realized using an additional unitary prior to computational basis readout, whose effect can similarly be analyzed as part of the general encoding $\mathcal{U}(u;\bm{\theta})$.


Note that for this ansatz, the choice $J=0~(\mbox{mod } \pi)$ yields either $\mathcal{W} = \hat{I}$ or $\hat{\sigma}^z \otimes \hat{\sigma}^z$, both of which ensure $\hat{\rho}(u)$ is a product state and measured features are simply products of uncorrelated individual qubit observables -- equivalent to a noisy classical system. Starting from this \textit{product system} (PS), tuning the coupling $J\neq 0~(\mbox{mod } \pi)$ provides a controllable parameter to realize a \rev{\textit{quantum correlated system} (CS), for which the $2^L$-dimensional multinomial distribution $\bm{x}(u)$ cannot be represented as a tensor product of $L$ marginal binomial distributions on each qubit. In general, such non-product systems intuitively result in $u$-dependent quantum states which exhibit entanglement and can potentially be more difficult to describe classically.}
This control enables us to address a natural question regarding EC of quantum systems under finite $\NS$: what is the dependence of EC and realizable eigentasks on $J$, and hence on quantum correlations? 

\rev{\subsection{Expressive Capacity of Quantum Circuits}}

To perform the capacity analysis, one must extract measured features from the quantum system as the input $u$ is varied, as exemplified in Fig.\,\ref{fig:Genc1}(a) for the IBMQ \emph{ibmq\_perth} device.  For comparison, we also show ideal-device simulations (unitary evolution, no device noise), where slight deviations are observed. The agreement with experimental results is improved when the effects of gate errors, readout errors, and qubit relaxation are included, hereafter referred to as ``device noise'' simulations, highlighting both the non-negligible role of device nonidealities, and that our analysis incorporates them.

The measured features under finite $\NS$ are used to estimate the Gram and covariance matrices (see detailed techniques in Appendix \ref{app:Spectral_finite_statistics}), and to therefore solve the eigenproblem Eq.\,(\ref{eq:eigenprob}) for NSR eigenvalues $\{\beta_k^2\}$. Typical NSR spectra computed for a random encoding (i.e.~set of rotation parameters) on the device are shown in Fig.\,\ref{fig:Genc1}(b), for $J=0$ (PS) and $J=\pi/2$ \rev{(CS)}, together with corresponding spectra from device noise simulations, with which they agree well. We note that at lower $k$, the device NSR eigenvalues are larger than those from ideal simulations, and at larger $k$ deviate from the direct exponential increase (with order) seen in ideal simulations. Both these effects are captured by device noise simulations as well and can therefore be attributed to device errors and dissipation. The NSR spectra therefore can serve as an effective diagnostic tool for quantum processors and encoding schemes. 


The NSR spectra can be used to directly compute the EC of the corresponding quantum device for finite $\NS$, via Eq.\,(\ref{eq:EC}). Practically, at a given $\NS$ only NSR eigenvalues $\beta_k^2 \lesssim \NS$ contribute substantially to the EC. An NSR spectrum with a flatter slope therefore has more NSR eigenvalues below $\NS$, which gives rise to a higher capacity. Fig.\,\ref{fig:Genc1}(b) shows that the CS generally exhibits an NSR spectrum with a flatter slope than the PS, yielding a larger capacity for function approximation across all sampled $\NS$.

To more precisely quantify the role of quantum correlations in EC, we introduce the \textit{expected total correlation} (ETC) of the measured state over the input domain of $u$ \cite{Vedral2002, Modi2010},
\begin{align}
   \bar{\mathcal{T}} = \mathbb{E}_u \! \left[ \sum_{l = 1}^L \mathrm{S} ( \hat{\rho}_l^{M} (u) ) - \mathrm{S} (\hat{\rho}^{M} (u) ) \right], 
\end{align}
where $\hat{\rho}^{M} (u) \equiv \sum_{k} \hat{\rho}_{kk}(u) \ket{\bm{b}_k}\!\bra{\bm{b}_k}$ is the post-measured state, $\mathrm{S}(\cdot)$ is the von Neumann entropy~(see Appendix \ref{app:QCM}),
and $\hat{\rho}_l = \mathrm{Tr}_{[L] \backslash \{ l \}} \{ \hat{\rho} \}$ is the reduced density matrix. \rev{Therefore, non-zero ETC signals the generation of quantum states over the input domain $u$ that on average have nontrivial correlations amongst their constituents, including for example pure many-body states that are entangled.}
We now compute EC and ETC using $S=2^{14}$ in Fig.\,\ref{fig:Genc1}(c) as a function of $J$, for the same random encoding considered above on the device. We note that the experimental results show excellent agreement in both cases with the corresponding device noise simulation; we also show average EC at $S=2^{14}$ and ETC across 8 random encodings in both ideal and device noise simulations.  The influence of individual encodings, i.e., random rotation parameters, is seen to manifest as small deviations from the overall EC trend governed by global hyperparameters, such as $J$ here (or $L$ in Appendix Fig.\,\ref{fig:Qubit_Scaling}).  This justifies our choice of random circuits to evaluate the overall capacity of an ansatz for learning.

We note that product states by definition have $\bar{\mathcal{T}}=0$~\cite{nielsen2002quantum}; this is seen in ideal simulations for $J=0~(\mbox{mod } \pi)$. However, the actual device retains a small amount of correlation at this operating point, which is reproduced by device noise simulations. This can be attributed to gate or measurement errors as well as cross-talk, the latter being especially relevant for the transmon-based IBMQ platform with a parasitic always-on ZZ coupling \cite{sheldon_procedure_2016}. 
With increasing $J$, $\bar{\mathcal{T}}$ increases and peaks around $J \approx \pi/2~(\mbox{mod } \pi)$; interestingly, $\EC$ also peaks for the same coupling range. From the analogous plot of EC, we clearly see that at finite $S$, increased ETC appears directly correlated with higher EC. We have observed very similar behaviour using completely different quantum system models~(see Appendix Fig.\,\ref{fig:NSR_Tilde_Renormalization}~\cite{Giovannetti2006, martinez2021dynamical}). This indicates the utility of enhancing quantum correlations as a means of improving the general expressive capability of quantum systems.

\rev{  We caution that this connection between measurement correlations and EC is an observed trend, rather than a law derived from first principles.  One can come up with contrived situations where increasing correlation has no effect on EC: for example, appending a layer of CNOT gates directly prior to measurement will generally increase the ideal ETC of any ansatz. For measured features however this amounts to a simple shuffling of labels  $x_k(\UI)\leftrightarrow x_{k'}(\UI)$, thus yielding the same NSR spectrum and EC. The input, quantum-state, and feature mapping ultimately governs EC: only increases in correlation that also increase the complexity of the measured features' $u$-dependence (as achieved via the intermediate $\mathcal{W}$ gates here) are beneficial from the perspective of information processing.
}

As a final important point, note that at finite $\NS$, even with increased quantum correlations, the maximum EC is still substantially lower than the upper bound of $K=64$. This remains true even for ideal simulations, and over several random encodings, so the underperformance cannot be attributed to device noise or poor ansatz choice respectively.
It is worth emphasizing that the impact of device noise is captured in the small EC gap between the ideal and noise simulation curves, with the remainder of the reduction from $K=64$ attributable to QSN alone.
These results clearly indicate that the resulting sampling noise at finite $\NS$ is the fundamental limitation for QML applications on this particular IBM device, rather than other types of noise sources and errors. 

\rev{\subsection{A Robust Approach to Learning}}
\label{sec:eigentasklearning}

While we have demonstrated the EC as an efficiently-computable metric of general expressive capability of a noisy quantum system, some important practical questions arise. First, does the general EC metric have implications for practical performance on \textit{specific} QML tasks? Secondly, given the limiting -- and unavoidable -- nature of correlated sampling noise, does the EC provide any insights on optimal learning using a particular noisy quantum system and the associated encoding?

\begin{figure}
    \centering
    \includegraphics[scale=1.0]{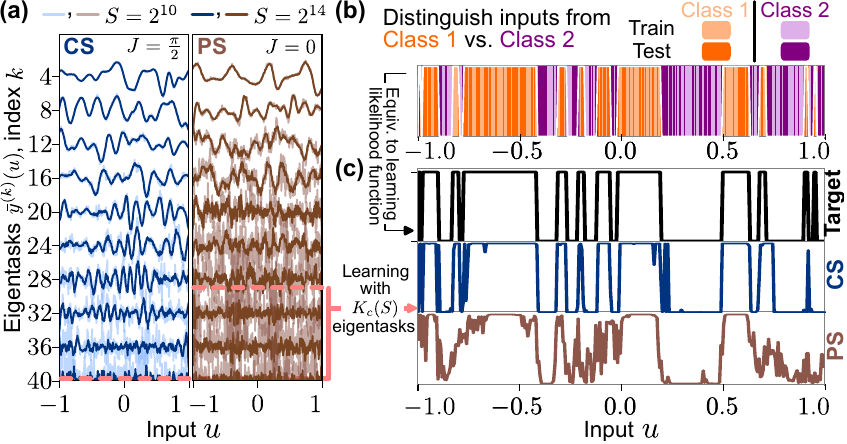}
    \caption{
    (a) Device eigentasks for \rev{correlated system (CS, left)} and \rev{product system (PS, right)}, constructed from noisy features at $S=2^{10}$ and $S=2^{14}$.
    (b) Classification demonstration on IBMQ Perth. Binary distributions to be classified over the input domain are shown. (c) The classification task can be cast as learning the likelihood function separating the two distributions; this target function is shown in the upper panel. Lower panels show the learned estimate of this target \rev{based on the $N_{\rm train}=150$ points shown in (b), using only $K_c(\NS)$ eigentasks for $\NS = 2^{14}$; this cutoff is indicated by the dashed red lines. For the \rev{correlated} system $K_c(S) = 40$, while for the product system $K_c(S) = 29$.}
    }
    \label{fig:Genc2}
\end{figure}

Our formulation addresses both these important questions naturally, as we now discuss. {Recall that} beyond being a simple figure of merit, the EC is precisely the sum of capacities to approximate a particular set of orthogonal functions native to the given noisy quantum system: the eigentasks. {Furthermore}, these eigentasks 
$\bar{y}^{(k)}(u)$ can be directly estimated from a noisy quantum system via the generalized eigenvectors $\{\bm{r}^{(k)}\}$, and are ordered by their associated NSR $\{\beta_k^2\}$. In Fig.\,\ref{fig:Genc2}(a) show a selection of estimated eigentasks \rev{from the device} for the CS $(\rev{J=\pi/2})$ and PS $(J=0)$ encodings of Fig.\,\ref{fig:Genc1}(b). For both systems, the increase in noise with eigentask order is apparent when comparing two sampling values, $\NS=2^{10}$ and $\NS=2^{14}$. Furthermore, for any order $k$, eigentasks for the PS are visibly noisier than the CS; this is consistent with NSR eigenvalues for PS being larger than those for CS (Fig.\,\ref{fig:Genc1}(b)). The higher expressive capacity of the CS can be interpreted the ability to accurately resolve more eigentasks at fixed $S$.


The resolvable eigentasks of a finitely-sampled quantum system are intimately related to its performance at specific QML applications. To demonstrate this result, we consider a concrete application: a binary classification task that is not linearly-separable. The domain $u \in [-1,1]$ over which EC was evaluated is separated into two classes, as depicted in Fig.\,\ref{fig:Genc2}(b). A selection of $N_{\rm train}=150$ total samples -- with equal numbers from each class -- are input to the IBMQ device, and 
eigentasks $\{\bar{y}^{(k)}(u^{(n)})\}_{K_{\rm L}}$ are estimated using $S=2^{14}$ shots. A linear estimator applied to this set of eigentasks is then trained using logistic regression to learn the class label associated with each input. Finally, the trained IBMQ device is used to predict class labels of $N_{\rm test}=150$ distinct input samples for testing. 
\rev{
Note that we use the \textit{random} circuits of the previous section to draw more direct comparisons between EC and task performance.  By training only external weights instead of internal parameters $\boldsymbol{\theta}$ we are employing the framework of Reservoir Computing \cite{fujii_harnessing_2017, dambre_information_2012}, which allows one to avoid the computational overhead and difficulty associated with training quantum systems while still achieving comparable performance \cite{chen_temporal_2020, Kalfus_2022, wright_capacity_2019, garcia-beni_scalable_2022}.
}

This task can equivalently be cast as one of learning the likelihood function that discriminates the two input distributions, shown in Fig.\,\ref{fig:Genc2}(c), with minimum error. The set of up to $K_{\rm L}$ eigentasks $\bar{y}^{(k)}(u)$, where $K_{\rm L} \leq K$, serves as the native orthonormal basis of readout features used to approximate \textit{any} target function using the quantum system. Importantly, the basis is \textit{ordered}, with eigentasks at higher $k$ contributing more noise, as dictated by the NSR eigenvalues $\beta_k^2$. In particular, at any level of sampling $\NS$, there exists an eigentask order $K_c(\NS)$ after which the NSR $\beta_k^2/\NS$ first drops below unity: $K_{c}(S) = \max_k\{\beta_{k}^2 < \NS\}$. Heuristically, including eigentasks $k > K_c(\NS)$ should contribute more `noise' to the function approximation task than `signal'. In Fig.\,\ref{fig:Genc2}(c), we plot the learned estimates of the likelihood function using $K_{\rm L} = K_c(\NS)$ eigentasks for both the CS and PS. First, we note that $K_c$ is lower for the PS than the \rev{CS}; the former has fewer resolvable eigentasks at a given $\NS$. This limitation on resolvable features limits function approximation capacity: the learned estimate of the likelihood function using $K_c$ eigentasks is visibly worse for the PS than the \rev{CS}. 




\begin{figure}[t]
    \centering
    \includegraphics[width=\columnwidth]{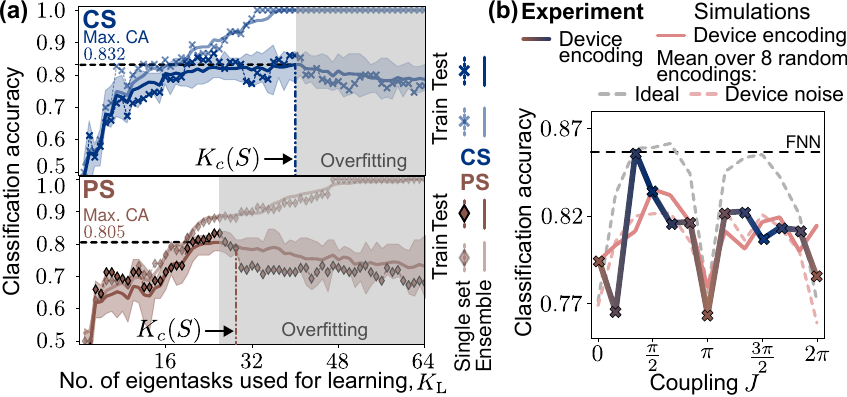}
    \caption{
    (a) Training (light) and testing (dark) accuracy for the 
    \rev{device encodings of Fig.\,\protect\ref{fig:Genc3}(a), as a function of the number of eigentasks used to approximate the target function.  Markers indicate performance on the dataset shown in Fig.\,\protect\ref{fig:Genc3}(b), and solid lines are the average over $10$ random selections of training and test sets.  The shaded region denotes the maximum and minimum test accuracy observed.
    The optimal test set performance is found near the noise-to-signal cutoff $K_c(S=2^{14})$ (dash-dotted lines) informed by the quantum system's noise-to-signal spectra.
    } (b) Testing set classification accuracy as a function of $J$ for our optimal learning method. \rev{In all cases, the average performance over the $10$ task permutations is reported, using $K_c(S=2^{14})$.   Markers indicate device results for the chosen encoding, and the corresponding simulation is shown in solid peach.  Dashed peach shows the average of these results over the $8$ device noise simulation encodings, and dashed grey the ideal simulation performance in the $S\to\infty$ limit, where all $K=64$ features are used.  The horizontal line denotes the performance of a software neural network with $K_{\rm L}=64$ nodes (and $1153 \gg K_c$ trained parameters) for comparison.
    }
    }
    \label{fig:Genc3}
\end{figure}


In this way, higher EC allows noisy quantum systems to better approximate more functions, which translates to improved learning performance -- this result is explored systemically in Fig.\,\ref{fig:Genc3}(b). Of course, it is natural to ask whether using $K_c(\NS)\leq K$ eigentasks is optimal: exactly this question is investigated in Fig.\,\ref{fig:Genc3}(a), where we plot the training and test accuracy of both device encodings as a function of the number of measured eigentasks $K_{\rm L}$. \rev{The performance on the specific training and test set shown in  Fig.\,\ref{fig:Genc2}(b) is indicated with markers, and solid lines indicate the average performance over $10$ distinct divisions of the data into training and test sets. This permutation of the learning task is a standard technique to optimize hyperparameters in ML, and is done here to eliminate the sensitivity of these results to the choice of training set.} First note that in all cases, using all eigentasks ($K_{\rm L} = K$) -- or equivalently all measured features $\{\bar{\bm{X}}\}$ -- leads to far lower test accuracy than is found in training.  The observed deviation is a distinct signature of overfitting: the optimized estimator learns noise in the training set (comprised of noisy eigentask estimates $\bar{y}^{(k)}(u^{(n)})$), and thus loses generalizability {to unseen samples} in testing.  

Improvements in model training performance with added features are only meaningful insofar as they also lead to better performance on new data: in both encodings we see test set classification accuracy peaks near $K_c(S)$. This is particularly clear for the averaged results, but even for individual datasets the test accuracy at $K_c(S)$ is within $\approx\!2\%$ of its maximum, thus confirming our heuristic reasoning that eigentasks beyond this order, with an NSR $<\!\!1$, hinder learning. The eigentask-learning approach naturally allows one to decompose the outputs from quantum measurements into a compressed basis with known noise properties, and then select the set of these which exactly captures the resolvable information at a given $S$.  This robust approach to learning enabled by the capacity analysis maximizes the ability of a noisy quantum system to approximate functions without overfitting to noise, in this case fundamental QSN.


Finally, Fig.\,\ref{fig:Genc3}(b) shows the classification accuracy for this device encoding as $J$ is varied, where following the above approach, the optimal $K_c(S)$ set of eigentasks are used for each encoding. 
\rev{
We also show the performance of a similar-scale ($K_{\rm L}=64$ node) software neural network and ideal simulations in the $S\to\infty$ limit ($K_c(\infty)=64$) for comparison.  Note that only these infinite-shot results approach the classical neural network, with QSN imposing a significant performance penalty even for $J \approx \pi/2~(\mbox{mod } \pi)$.
}
We highlight the striking similarity with Fig.\,\ref{fig:Genc1}(c): encodings with larger quantum correlations and thus higher expressive capacity will perform generically better on learning tasks in the presence of noise, because they generate a larger set of eigentasks that can be resolved at a given sampling $\NS$. Expressive Capacity is a priori unaware of the specific problem considered here; this example thus emphasizes its power as a general metric predictive of performance on arbitrary tasks.

\section{Discussion} 

We have developed a straightforward approach to quantify the expressive capacity of any quantum system in the presence of fundamental sampling noise. Our analysis is built upon an underlying framework that determines the native function set that can be most robustly realized by a finitely-sampled quantum system: its eigentasks. We use this framework to introduce a methodology for optimal learning using noisy quantum systems, which centers around identifying the minimal number of eigentasks required for a given learning task. The resulting learning methodology is resource-efficient and robust to overfitting. We demonstrate that eigentasks can be efficiently estimated from experiments on real devices using a limited number of training points and finite shots. We also demonstrate across two distinct qubit-based ans\"atze that the presence of measured quantum correlations enhances expressive capacity. Our work has direct application to the design of circuits for learning with qubit-based systems. In particular, we propose the optimization of expressive capacity as a meaningful goal for the design of quantum circuits with finite measurement resources.

\section*{Acknowledgement}
We would like to thank  Ronen Eldan, Daniel Gauthier, Michael Hatridge, Benjamin Lienhard, Peter McMachon, Sridhar Prabhu, Shyam Shankar, Francesco Tacchino, Logan Wright for stimulating discussions about the work that went into this manuscript. This research was developed with funding from the DARPA contract HR00112190072, AFOSR award FA9550-20-1-0177, and AFOSR MURI award FA9550-22-1-0203. The views, opinions, and findings expressed are solely the authors and not the U.S. government. 

\appendix

\bibliography{bibtex}


\begin{widetext}

\newpage


\section{Table of Symbols and Abbreviations}
\label{app:table}


\begin{table}[htb]
\begin{tabular}{cp{0.66\textwidth}}

\toprule
    \multicolumn{2}{c}
    {\textbf{Abbreviations}}  \\
    \hline
    NISQ & Noisy Intermediate Scale Quantum \\
    (Q)ML & (Quantum) Machine Learning \\
    QSN & Quantum Sampling Noise \\
    VQC & Variational Quantum Circuits \\
    PS & Product System \\
    CS & Correlated System \\
    EC  & Total Expressive Capacity, $C_T$ \\
    NSR & Noise-to-signal ratio \\
    ETC  & Expected Total Correlation, $\bar{\mathcal{T}}$ \\
    \hline
    \multicolumn{2}{c}
    {\textbf{Symbols and notation}} \\
    \hline    
    $S$ & Number of shots \\
    $N$ & Number of inputs \\
    $L$ & Number of qubits \\
    $K$ & Number of measured features; $K= 2^L$ for computational-basis projective measurement \\
    $\UI$ & Input \\
    $\bm{\theta}$ & Quantum system parameters \\
    $\hat{\rho}$ & Generated quantum state \\
    $\hat{M}_k$ & POVM elements, $\equiv  \ketbra{\boldsymbol{b}_k }$ for computational-basis projective measurement\\
    $\bm{W}$ & General output weights \\
    $\bm{w}$ & \textit{Learned} Optimal output weights for finite-$S$ features $\{\bar{X}_k\}$ \\
    $\mathscr{L}$ & Loss function \\
    ${\bm b}_k$ & Computational basis eigenstate label\\
    $k^{(s)}$ & Measurement outcome for shot $s$ \\
    $x_k$ & Expected features, ${\rm Tr}\{\hat{M}_k \hat{\rho} \}$ \\
    $\bar{X}_k$ & Empirical observed feature, $(1/S) \sum_s \delta(k^{(s)}, k)$ \\
    $\zeta_{k}$ & Noise component of $\bar{X}_{k}$ \\
    $\gr$ & Gram matrix of expected features $\{x_k\}$ \\
    $\ci$ & Expected covariance matrix of random variable $X^{(s)}_k(\UI)$ over input distribution \\
    $\beta_k^2$ & Eigen-NSR associated with eigentask $k$ \\
    $y^{(k)}$ & Eigentask, $\sum_{k'} r_{k'}^{(k)} x_{k'}$ \\
    $\bm{r}^{(k)}$ & linear combination of expected features $\{x_{k'}\}$ forming  $y^{(k)}$ \\
    $\bar{y}^{(k)}$ & Finite-$S$ estimate of eigentask, $\sum_{k'} r_{k'}^{(k)} \bar{X}_{k'}$ \\
    $\hat{\rho}^{M}$ & diagonal post-measurement state, $\sum_{k}\hat{\rho}_{kk}(u) \ket{\bm{b}_k}\!\bra{\bm{b}_k}$\\
    \hspace{8mm} $K_c(\NS)$ \hspace{8mm} & Cutoff index where $\beta_k^2$ approaches $\NS$, $\max_k\{\beta_{k}^2 < \NS\}$\\
    \hline
\end{tabular}
\caption{Table of notations used in main text}
\end{table}


\section{Feature maps using quantum systems}
\label{sec:NISQRC_Architecture_Detail}

\label{DetailsEncodings}

In the main text, we introduce the idea of encoding inputs into the state of a quantum system via a parameterized quantum channel, reproduced below:
\begin{align}
    \hat{\rho}(\bm{u};\bm{\theta}) = \mathcal{U}(\bm{u};\bm{\theta})\hat{\rho}_0,
    \label{appeq:quantumChannel}
\end{align}
one then measures this state to approximate desired functions of the input. 
\rev{Figure \ref{fig:Schematic_4Inputs} gives a simple example of this mapping from classical inputs $\UI$ (here in a 2D compact domain) to a quantum state generated by a $\bm{u}$-dependent encoding, and finally to the measured features in a $2$-qubit system undergoing commuting local measurements in the computational basis. The measurement outcomes are therefore bitstrings, of which there are $K= 2^L = 4$, namely: $\bm{b}_k \in \{00,01,10,11\}$. A given shot will yield one of these possible bitstrings.}

\rev{On the right we plot samples of features $\bar{\bm{X}}_k$ constructed with different numbers of shots $S$. As expressed in Eq.\,\eqref{eq:xbar}, the noise and thus variance in the distribution of samples scales with $S$. As $S\to \infty$ this distribution thus collapses to a single deterministic point, the corresponding quantum probability $\bm{x}(\UI)$.  It is also evident from this plot that the shape and orientation of these clusters depends on the underlying quantum state $\hat{\rho}(\bm{u};\bm{\theta})$ and associated probabilities $\bm{x}(\UI)$ via Eq.\,\eqref{eq:cov}.  In the remainder of this section, we will consider more complex quantum models, such that they generate mappings which can be useful for learning.
}

To describe these models, we begin by first limiting to 1-D inputs $u$ as analyzed in the main text; generalizations to multi-dimensional inputs $\bm{u}$ are straightforward. Then, we write Eq.\,(\ref{appeq:quantumChannel}) in the form
\begin{align}
    \hat{\rho}(u;\bm{\theta}) = \hat{U}(u; \boldsymbol{\theta}) \hat{\rho}_0 \hat{U}^{\dagger}(u; \boldsymbol{\theta})
\end{align}
In the main text, we have considered a model for dynamics of an $L$-qubit quantum system that is natively implementable on modern quantum computing platforms: namely an ansatz of quantum circuits with single and two-qubit gates. We refer to this encoding as the \textit{circuit ansatz} (or \textit{C-ansatz} for short) for which the operator $\hat{U}(u; \boldsymbol{\theta})$ takes the precise form
\begin{align}
    \hat{U}(u; \boldsymbol{\theta}) = \left[ \mathcal{R}_{x} \!\left(\frac{\bm{\theta}^x}{2}\right) \mathcal{W}(J) \mathcal{R}_{z} \!\left(\bm{\theta}^z +\bm{\theta}^I u \right) \mathcal{R}_{x} \!\left(\frac{\bm{\theta}^x}{2} \right)
    \right]^{\tau}~~~~~~~\textit{(C-ansatz)}
    \label{appeq:cansatz}
\end{align}
\begin{figure}
    \centering
    \includegraphics[width = 0.8\columnwidth]{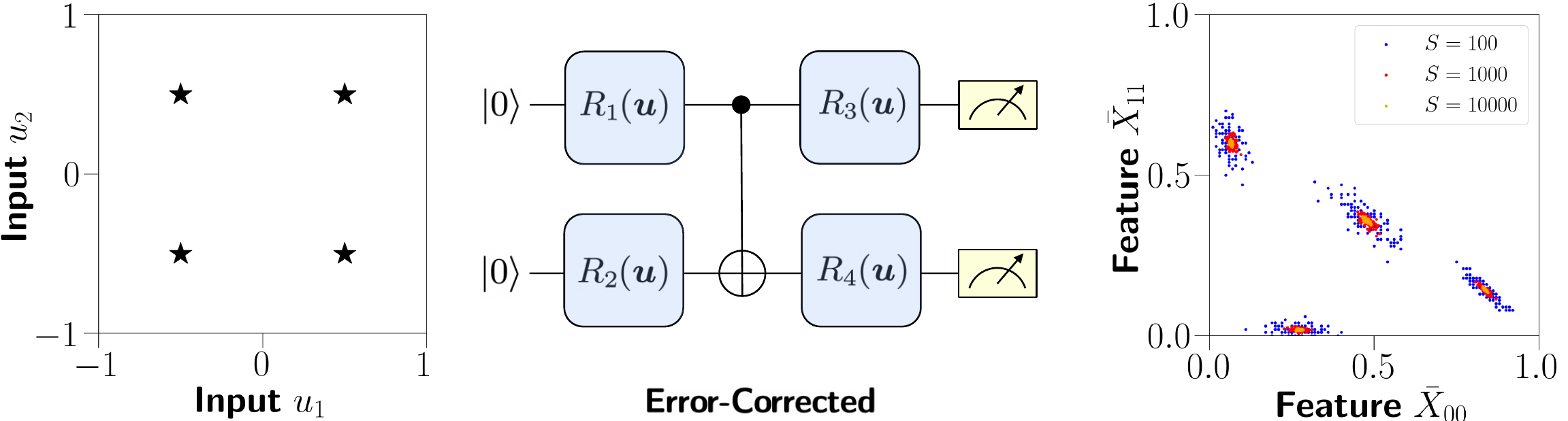}
    \caption{
    \rev{Schematic of a simple $L=2$ qubit circuit, comprised of a CNOT gate sanwiched by input-dependent local $x$-rotation gates $\{R_i(\UI)\}$.  Different $2$D inputs shown on the left are mapped to the finite-$S$ feature space on the right via this circuit. Specifically, a $2$D slice ($\bar{X}_{00}$ and $\bar{X}_{11}$) of the $4$D feature space is shown. Each point represents an individual sample or experiment, i.e. an output constructed with $S<\infty$ shots via Eq.\,\protect\eqref{eq:Xsum}. Distinct values of $S=10^2,10^3,10^4$ are shown in different colors (blue, red, green). For each input $\UI$ and shots $S$, the simulation is conducted for $100$ repetition. 
    }}
    \label{fig:Schematic_4Inputs}
\end{figure}

For completeness, we recall that $ \mathcal{R}_{x/z}$ are Pauli-rotations applied qubit-wise, e.g.\,$ \mathcal{R}_{z} = \prod_{l} R_{z}({\theta}^z_l +{\theta}^I _l u) $, while the coupling gate acts between physically connected qubits in the device and can be written as $\mathcal{W}(J) = \prod_{\langle l, l' \rangle} \mathrm{exp}\{- i \frac{J}{2} \hat{\sigma}^z_{l} \hat{\sigma}^z_{l'}\} $. We emphasize here again that $\tau \in \mathbb{N}^+$ is an integer, representing the number of repeated blocks in the C-ansatz encoding. We note that the actual operations implemented on IBMQ processors also include dynamics due to noise, gate, and measurement errors, and thus must be represented as a general quantum channel as in Eq.\,\eqref{appeq:quantumChannel}. As discussed in the main text, the EC of a quantum system can be computed in the presence of these more general dynamics, and is sensitive to the limitations introduced by them.

An alternative ansatz which we analyze in this SI, is where the operator $\hat{U}(u; \boldsymbol{\theta})$ describes continuous Hamiltonian dynamics. This ansatz is relevant to computation with general quantum devices, such as quantum annealers and more generally quantum simulators. In this case, which we refer to as the \textit{Hamiltonian ansatz} (or \textit{H-ansatz} for short), 
\begin{align}
    \hat{U}(u; \boldsymbol{\theta}) = {\rm exp}\{-i\hat{H}(u)t \},~\hat{H}(u) = \hat{H}_0 + u \cdot \hat{H}_1~~~~~~~\textit{(H-ansatz)}
    \label{eq:Hu=H0+uH1}
\end{align}
Here $t$ is a continuous parameter defining the evolution time; and $\hat{H}_0 = \sum^L_{l, l'} J_{\langle l,l' \rangle} \hat{\sigma}^z_l \hat{\sigma}^z_{l'} + \sum^L_{l=1} h^x_{l} \hat{\sigma}^x_l  + \sum^L_{l=1} h^z_{l} \hat{\sigma}^z_l$ and $\hat{H}_1 = \sum^L_{l=1} h^I_{l} \hat{\sigma}^z_l$. The transverse $x$-field strength $h^x_{l} = \bar{h}^x + \varepsilon^x_{l}$ and longitudinal $z$-drive strength $h^{z,I}_{l} = \bar{h}^{z,I} + \varepsilon^{z,I}_{l}$ are all randomly chosen and held fixed for a given realization of the quantum system,
\begin{align}
    \varepsilon^{x,z,I}_{l} \sim h^{x,z,I}_{\mathrm{rms}}~\mathcal{N}(0, 1),
\end{align}
where $\mathcal{N}(0,1)$ defines the standard normal distribution with zero mean and unit variance. We consider nearest-neighbor interactions $J_{l,l'}$, which can be constant $J_{l,l'} \equiv J$, or drawn from $J_{l,l'} \sim \mathrm{Unif}[0, J_{\rm max}]$, where $\mathrm{Unif}[a,b]$ is a uniform distribution with non-zero density within $[a,b]$. 

As an aside, we note that the C-ansatz quantum channel described by Eq.\,(\ref{appeq:cansatz}) can be considered a Trotterization-inspired implementation of the H-ansatz in Eq.\,(\ref{eq:Hu=H0+uH1}). In particular, if we set $\theta^{x/z/I} = h^{x/z/I} \Delta\cdot\tau$, where $t=\Delta \cdot \tau$, and consider the limit $\Delta \to 0$ while keeping $t$ fixed, Eq.\,(\ref{appeq:cansatz}) corresponds to a Trotterized implementation of Eq.\,(\ref{eq:Hu=H0+uH1}). This correspondence is chosen for practical reasons, but is not necessary in our analysis.


\RestyleAlgo{ruled}
\SetKwComment{Comment}{/* }{ */}
\SetKwInOut{input}{Input}
\SetKwInOut{output}{Output}
\SetKwFor{For}{For}{}{EndFor}
\SetKwFor{If}{If}{}{EndIf}
\begin{algorithm}[t]
    \caption{Measured features in the \textbf{probability} representation}
    \label{alg:arc_prob}
    \input{$u \in [-1, +1]$}
    \output{$\bar{\boldsymbol{X}}(u)$, which approximates $x_k(u) := \mathrm{Tr}\left\{\hat{\rho}(u) \ket{\boldsymbol{b}_k} \! \bra{\boldsymbol{b}_k}\right\}$}
    \For{$s \gets 1$ to $\NS$}{
        Initialize overall state $\hat{\rho}_{0} \gets \ket{0}\bra{0}^{\otimes \NQ}$\;
        Evolve under quantum channel $\mathcal{U}(u)$: $\hat{\rho}(u) \gets \mathcal{U}(u) \hat{\rho}_0$\;
        Measure all $\NQ$ qubits: $\boldsymbol{b}^{(s)}(u) \gets \boldsymbol{b}_k =  \left(b_{k,1}, b_{k,2} \cdots, b_{k,L}\right) \in \{0, 1\}^{\NQ}$\;
    }
    \For{$k \gets 0$ to $K-1$}{
        Take the ensemble averages as readout features:\\
        \hspace{5mm} $\bar{X}_k(u) \gets \frac{1}{\NS}\sum_{s=1}^{\NS} \delta (\boldsymbol{b}_k, \boldsymbol{b}^{(s)}(u))$ \Comment*[r]{Notice $x_k(u) := \mathrm{Tr}\left\{\hat{\rho}(u) \ket{\boldsymbol{b}_k} \! \bra{\boldsymbol{b}_k}\right\} = \lim_{\NS \to \infty} \bar{X}_k(u)$}
    }
\end{algorithm}

\begin{algorithm}[t]
    \caption{Training of output weights}\label{alg:training}
    \input{$\{ u^{(1)}, \cdots, u^{(N)} \} \in [-1, +1]^N$}
    \output{$\widetilde{\boldsymbol{w}}_N$, such that $y = \widetilde{\boldsymbol{w}}_N \cdot \bar{\boldsymbol{X}}(u)$ can approximate $f(u)$}
    \For{$n \gets 1$ to $N$}{
        Generate features $\bar{\boldsymbol{X}}(u^{(n)})$  through Algorithm 1
    }
    Collect the features into a regression matrix $\regmat \in \mathbb{R}^{N \times K}$\;
    Compute empirical Gram matrix $\bar{\gr} \gets \frac{1}{N} \regmat^T \regmat$ \Comment*[r]{For finite $S$, $\lim_{N \to \infty} \bar{\gr} = \tilde{\gr} := \gr + \frac{1}{S} \ci$}
    Compute target vector $\boldsymbol{Y} \gets \left(f(u^{(1)}), \cdots, f(u^{(N)}) \right)^T$\;
    $\widetilde{\boldsymbol{w}}_N \gets (\regmat^T \regmat)^{-1} \regmat^T \boldsymbol{Y} $ \Comment*[r]{For finite $S$, $\lim_{N \to \infty} \widetilde{\boldsymbol{w}}_N = \widetilde{\boldsymbol{w}} :=$ Eq.\,(\ref{eq:wopt})}
\end{algorithm}

\section{Information capacity with quantum sampling noise}
\label{sec:Information_capacity_saturation}

\subsection{Definition of capacity for quantum systems with sampling noise}
A universal function approximation theorem (which will be formally stated in Appendix \ref{sec:Single_step_QRC}), as a basic requirement of most neural network models, can be made concrete by defining a metric to quantify how well a given quantum system (or \textit{any} dynamical system) approximates general functions. Suppose an arbitrary probability distribution $p(u)$ for a random (scalar) variable $u$ defined in $D \subseteq \mathbb{R}$. This naturally defines a function space $L^2_p (D)$ containing all functions $f:D \to \mathbb{R}$ with $\int f^2(u) p(u) \dd u < \infty$. The space is equipped with the inner product structure $\langle f_1, f_2 \rangle_{p} = \int f_1(u) f_2(u) p(u) \dd u$. A standard way to check the ability of fitting nonlinear functions by a physical system is the \textit{information processing capacity}~\cite{dambre_information_2012},
\begin{equation}
    C [f_{\ell}] = 1 - \min_{\boldsymbol{W}_\ell \in \mathbb{R}^K}  \frac{\int \left( \sum_{k = 0}^{K-1} W_{\ell k} x_k (u) - f_{\ell} (u) \right)^2 p(u) \dd u}{\int f_{\ell} (u)^2 p(u)  \dd u},
\end{equation}
where functions $f_{\ell}(u)$ are orthogonal target functions $\langle f_{\ell}, f_{\ell'} \rangle_{p} = \int f_{\ell}(u) f_{\ell'}(u) p(u) \dd u = 0$ for $\ell \neq \ell'$. The \textit{total expressive capacity} is defined as $\EC \equiv \sum_{\ell=0}^{\infty} C[f_{\ell}]$, capturing the ability of what type of function the linear combination of physical system readout features can produce. Dambre \textit{et. al.}'s argument claims that the total capacity must be upper bounded by the number of features $\EC \leq K$.

While Dambre \textit{et. al.}'s result~\cite{dambre_information_2012} is quite general, it neglects the limitations due to noise in readout features, a fact that is unavoidable when using quantum systems in the presence of finite computational and measurement resources. It is generally accepted that the capacity is reduced in the presence of additive noise, but there are no general results on how to quantify that reduction. This is our goal here, to arrive at an exact result for capacity reduction under well-defined conditions.  

In this section, we will focus on the impact of fundamental quantum readout noise, or quantum sampling noise (QSN), on this upper bound under finite sampling $\NS$. Given $u$ and $\NS$, the quantum readout features $\bar{X}_k (u) = \frac{1}{\NS} \sum_{s=1}^{\NS} \delta(k^{(s)} (u), k)$ are stochastic variables. The expectation vector and covariance matrix of $\bar{\bm{X}} (u)$ can be expressed in terms of $\hat{\rho} (u)$
\begin{align}
    \mathbb{E} [\bar{\boldsymbol{X}} (u)] & \equiv \boldsymbol{x}(u) = \mathrm{Tr}\{\hat{E_k} \hat{\rho}(u)\}, \label{eq:Ux}\\
    \mathrm{Cov} [\bar{\boldsymbol{X}} (u)] & \equiv \frac{1}{\NS} \cu(u) = \frac{1}{\NS} \left(\mathrm{diag}(\bm{x})-\bm{x}\bm{x}^T \right). \label{eq:Sigma}
\end{align}

To determine the optimal capacity to compute an arbitrary normalized function $f (u) = \sum_{j = 0}^{\infty} (\mathbf{Y})_j u^j$ using the noisy readout features $\bar{\bm{X}}(u)$ extracted from the quantum system, we need to find an optimal $\boldsymbol{W}$ such that
\begin{equation}
    C [f] = 1 - \frac{\min_{\boldsymbol{W}}  \int \left( \sum_{k = 0}^{K-1} W_k \bar{X}_k (u) - f (u) \right)^2 p(u) \dd u}{\int f^2 (u) p(u) \dd u} \label{eq:defIPC}
\end{equation}

By expanding the numerator of the right-hand side for a given, finite number of shots $\NS$, we find
\begin{align}
     & \int f^2 (u) p(u) \dd u - \int \left( \sum_{k = 0}^{K-1} W_k \bar{X}_k (u) - f (u) \right)^2\!\! p(u) \dd u \nonumber\\
     =~&\! - \sum_{k_1 = 0}^{K-1} \sum_{k_2 = 0}^{K-1} W_{k_1} W_{k_2}  \int \bar{X}_{k_1} (u) \bar{X}_{k_2} (u) p(u) \dd u + 2 \sum_{k = 0}^{K-1} W_k  \int \bar{X}_k (u) f (u) p(u) \dd u \nonumber\\
    \approx~&\! - \frac{1}{\NI} \sum_{k_1 = 0}^{K-1} \sum_{k_2 = 0}^{K-1} W_{k_1} W_{k_2}  \sum_{n = 1}^{\NI} \bar{X}_{k_1} (u^{(n)}) \bar{X}_{k_2} (u^{(n)}) + \frac{2}{\NI} \sum_{k = 0}^{K-1} W_k  \sum_{n = 1}^{\NI} \bar{X}_k (u^{(n)}) f (u^{(n)}) . 
\end{align}
where we have approximated the integral over the input domain by a finite sum in the limit of a large number of inputs $\NI$. Next, note that if $n \neq n'$, then $X_{k_1} (u^{(n)})$ and $X_{k_2} (u^{(n')})$ are independent random variables (though not necessarily identically distributed). The sums over $N$ on the right hand side are therefore sums of bounded independent random variables. In the limit of large $\NI \gg 1$, the deviation between stochastic realizations of these sums and their expectation values is exponentially suppressed, as determined by the Hoeffding inequality. Then, with large probability, the sums over $N$ may be replaced by their expectation values,
\begin{align}
    & \int f^2 (u) p(u) \dd u - \int \left( \sum_{k = 0}^{K-1} W_k \bar{X}_k (u) - f (u) \right)^2 \!\!p(u) \dd u \nonumber\\
    \approx~& - \frac{1}{\NI} \sum_{k_1 = 0}^{K-1} \sum_{k_2 = 0}^{K-1} W_{k_1} W_{k_2}  \sum_{n = 1}^{\NI} \mathbb{E} [\bar{X}_{k_1} (u^{(n)}) \bar{X}_{k_2} (u^{(n)})] + \frac{2}{\NI} \sum_{k = 0}^{K-1} W_{k}  \sum_{n = 1}^{\NI} \mathbb{E} [\bar{X}_k (u^{(n)}) f (u^{(n)})] \nonumber\\
    =~& - \frac{1}{\NI} \sum_{k_1 = 0}^{K-1} \sum_{k_2 = 0}^{K-1} W_{k_1} W_{k_2}  \sum_{n = 1}^{\NI} \left( x_{k_1} (u^{(n)}) x_{k_2} (u^{(n)}) + \frac{1}{\NS} \cu(u^{(n)})_{k_1 k_2} \right) + \frac{2}{\NI} \sum_{k = 0}^{K-1} W_k \sum_{n = 1}^{\NI} x_k (u^{(n)}) f (u^{(n)}) \nonumber\\
    \approx~& - \sum_{k_1 = 0}^{K-1} \sum_{k_2 = 0}^{K-1} W_{k_1} W_{k_2}  \int \left( x_{k_1} (u) x_{k_2} (u) + \frac{1}{\NS} \cu(u)_{k_1 k_2} \right) p(u) \dd u + 2 \sum_{k = 0}^{K-1} W_k  \int x_k (u) f (u) p(u) \dd u. \label{eq:Lambdatilde}
\end{align}
The first approximation above comes from the Hoeffding inequality, where terms that are dropped are proportional to $1/\sqrt{\NI}$. In going from the second to the third line, we have used Eq.\,(\ref{eq:Sigma}). The final expression is obtained by rewriting sums over $u$ as integrals, with an error proportional to $1/\sqrt{\NI}$ once more. Thus we can say the original integral in Eq.\,(\ref{eq:defIPC}) is approximately equal to Eq.\,(\ref{eq:Lambdatilde}) to $O (1/\sqrt{\NI})$. \rev{In the limit of a large number of input samples, $N \to \infty$, we conclude that all approximations can be replaced by exact equalities. }

\rev{The goal of the remaining part of this section is deducing a more compact generalized Rayleigh quotient form of functional capacity.} The dependence of readout features $x_k(u)$ on the input $u$ can always be written in the form of a Taylor expansion,
\begin{align}
    x_k (u) = \sum_{j = 0}^{\infty} (\mathbf{T})_{k j} u^j
    \label{eq:TMat}
\end{align}
where we define the \textit{transfer matrix} $\mathbf{T}(\bm{\theta})\equiv \mathbf{T} \in \mathbb{R}^{K \times \infty}$ that depends on the density matrix $\hat{\rho}(u)$, and in particular on parameters $\bm{\theta}$ characterizing the quantum system. 
The first term in Eq.\,(\ref{eq:Lambdatilde}) does not depend explicitly on the function $f(u)$ being constructed, and introduces quantities that are determined entirely by the response of the quantum system of interest to inputs over the entire domain of $u$. In particular, we introduce the \textit{Gram matrix} $\gr \in \mathbb{R}^{K \times K}$ as
\begin{align}
    (\gr)_{k_1k_2} &= \int x_{k_1}(u) x_{k_2} (u) p(u) \dd u  = \sum_{j_1 = 0}^{\infty} \sum_{j_2 = 0}^{\infty} (\mathbf{T})_{k_1 j_1} \left( \int u^{j_1 + j_2} p(u) \dd u \right) (\mathbf{T})_{k_2 j_2} \equiv (\mathbf{T}\gh\mathbf{T}^T)_{k_1k_2}
\end{align}
where in the second line we have also introduced the \textit{generalized Hilbert matrix} $\gh \in \mathbb{R}^{\infty \times \infty}$ as
\begin{equation}
    (\gh)_{j_1 j_2} = \int u^{j_1 + j_2} p(u) \dd u .
\end{equation}
Secondly, we introduce the noise matrix $\ci \in \mathbb{R}^{K \times K}$,
\begin{align}
    (\ci)_{k_1 k_2} & = \int \cu(u)_{k _1k_2}~p(u) \dd u = \int (\delta_{k_1 k_2} x_{k_1} (u) - x_{k_1}\!(u) x_{k_2}\!(u)) p(u) \dd u \equiv (\mathbf{D})_{k_1k_2}-(\gr)_{k_1k_2}
    \label{eq:V}
\end{align}
Here we have also introduced the \textit{second-order-moment} matrix $\mathbf{D} \in \mathbb{R}^{K \times K}$ such that $(\mathbf{D})_{k_1 k_2} = \delta_{k_1 k_2} \int x_{k_1} (u) p(u) \dd u$. Then, the noise matrix simply defines the covariance of readout features, and is therefore given by $\ci = \mathbf{D} - \gr$.
The second term in Eq.\,(\ref{eq:Lambdatilde}) depends on $f(u)$ and can be simplified using the $\gh$ matrix as well,
\begin{align} 
    \int x_{k} (u) f (u) p(u) \dd u & = \sum_{j_1 = 0}^{\infty} \sum_{j_2 = 0}^{\infty} (\mathbf{T})_{k j_1} \left( \int u^{j_1 + j_2} p(u) \dd u \right) (\mathbf{Y})_{j_2} = (\mathbf{T}\gh\mathbf{Y})_{k}.
\end{align}
With these definitions, Eq.\,(\ref{eq:defIPC}) can be compactly written in matrix form as a Tikhonov regularization problem: 
\begin{equation}
    C[f] =
    1 - \min_{\boldsymbol{W}} \left( \frac{ \left\| \gh^{\frac{1}{2}}\mathbf{T}^T \boldsymbol{W} - \gh^{\frac{1}{2}}\mathbf{Y} \right\|^2 + \frac{1}{\NS} \boldsymbol{W}^T \ci \boldsymbol{W}} {\mathbf{Y}^T\gh\mathbf{Y}} \right). \label{eq:Cf}
\end{equation}
The least-squares form ensures that the optimal value (argmin) $\boldsymbol{w}$ of $\boldsymbol{W}$ has closed form 
\begin{equation}
    \boldsymbol{w} = \left( \mathbf{T}\gh\mathbf{T}^T + \frac{1}{\NS} \ci \right)^{- 1} \mathbf{T}\gh\mathbf{Y}. \label{eq:wopt}
\end{equation}
Substituting $\bm{w}$ into the expression for $C$, we obtain the optimal capacity with which a function $f$ can be constructed, which takes the form of a \textit{generalized Rayleigh quotient}
\begin{equation}
    C[f] = \frac{\mathbf{Y}^T \gh\mathbf{T}^T \left( \gr + \frac{1}{\NS} \ci \right)^{- 1} \mathbf{T}\gh\mathbf{Y}}{\mathbf{Y}^T\gh\mathbf{Y}}. \label{eq:Rayleigh_quotient}
\end{equation}

\subsection{Eigentasks}

Eq.\,(\ref{eq:Rayleigh_quotient}) defines the optimal capacity of approximating an arbitrary function $f (u) = \sum_{j = 0}^{\infty} (\mathbf{Y})_j u^j$. We can therefore naturally ask which functions $f$ maximise this optimal capacity. To this end, we first note that the denominator of Eq.\,(\ref{eq:Rayleigh_quotient}) is simply a normalization factor that can be absorbed into the definition of the function $f(u)$ being approximated, without loss of generality. More precisely, we consider:
\begin{align}
    \langle f, f \rangle_p = 1 = \left( \gh^{\frac{1}{2}} \mathbf{Y} \right)^T \left( \gh^{\frac{1}{2}} \mathbf{Y} \right) = \mathbf{Y}^T\gh\mathbf{Y}.
\end{align}
Then, we can rewrite the optimal capacity from Eq.\,(\ref{eq:Rayleigh_quotient}) as
\begin{align}
    C[f] = \mathbf{Y}^T\gh^{\frac{1}{2}} \mathbf{Q} \gh^{\frac{1}{2}}\mathbf{Y}.
    \label{eq:Cp}
\end{align}
Here we have defined the matrix $\mathbf{Q} \in \mathbb{R}^{\infty \times \infty}$ as
\begin{align}
    \mathbf{Q} &= \mathbf{B}\left(\mathbf{I} + \frac{1}{\NS} \mathbf{R} \right)^{- 1} \!\!\! \mathbf{B}^{T}, \label{eq:Q} \\
    \mathbf{B} &= \gh^{\frac{1}{2}} \mathbf{T}^T\gr^{-\frac{1}{2}},
    \label{eq:B} \\
    \mathbf{R} & = \gr^{-\frac{1}{2}}\mathbf{V}\gr^{-\frac{1}{2}}
\end{align}
by introducing the matrix square root of $\gr^{\frac{1}{2}} \in \mathbb{R}^{K \times K}$, and $\mathbf{R}$ the \textit{noise-to-signal} matrix. The decomposition in Eq.\,(\ref{eq:Q}) may be verified by direct substitution into Eq.\,(\ref{eq:Cp}). The ability to calculate matrix powers and in particular the inverse of $\gr$ requires constraints on its rank, which we show are satisfied in Appendix \ref{app:Function-independence}.

We now consider the measure-independent part of the eigenvectors of $\mathbf{Q}$, indexed $\mathbf{Y}^{(k)}$, satisfying the standard eigenvalue problem:
\begin{align}
    \mathbf{Q} \gh^{\frac{1}{2}}\mathbf{Y}^{(k)} = C_k\gh^{\frac{1}{2}}\mathbf{Y}^{(k)}.
    \label{eq:eigQ}
\end{align}
where $k=0,\cdots,K-1$. From Eq.\,(\ref{eq:Cp}), it is clear that these eigenvectors have a particular meaning. Consider the function $y^{(k)}(u)$ defined by the eigenvector $\mathbf{Y}^{(k)}$, namely
\begin{align}
    y^{(k)}(u) = \sum_{j = 0}^{\infty} \mathbf{Y}^{(k)}_{j} u^j,
    \label{eq:fkdef}
\end{align}
which we will refer to from now on as \textit{eigentasks}. Suppose we wish to construct the function $y^{(k)}(u)$ using outputs obtained from the physical system defined by $\mathbf{Q}$ in the $\NS\to\infty$ limit (namely, with \textit{deterministic} outputs). At a first glance, before we dive into solving the eigenproblem Eq.(\ref{eq:eigQ}), we do not know any relationship between $y^{(k)}$ and $\bm{x}(u)$.The rest part of this subsection is aiming to  prove that $y^{(k)}$ must be a specific linear combination of features $\bm{x}(u)$. Then, the physical system's capacity for this construction is simply given by the corresponding eigenvalue $C_{k}$, as may be seen by substituting Eq.\,(\ref{eq:eigQ}) into Eq.\,(\ref{eq:Cp}). Formally, the $y^{(k)}(u)$ serves as the \textit{critical point} (or \textit{stationary point}) of the generalized Rayleigh quotient in Eq.\,(\ref{eq:Rayleigh_quotient}). Consequently, the function that is constructed with largest capacity then corresponds to the nontrivial eigenvector with largest eigenvalue. 

To obtain these eigentasks, we must solve the eigenproblem defined by Eq.\,(\ref{eq:eigQ}). Here, the representation of $\mathbf{Q}$ in Eq.\,(\ref{eq:Q}) becomes useful, as we will see that the eigensystem of $\mathbf{Q}$ is related closely to that of the noise-to-signal matrix $\mathbf{R}$. In particular, we first define the eigenproblem of $\mathbf{R}$,
\begin{align}
    \nsr \gr^{\frac{1}{2}}\bm{r}^{(k)} & = \beta_k^2 \gr^{\frac{1}{2}}\bm{r}^{(k)} \label{eq:Reigen}
\end{align}
with NSR eigenvalues $\beta_k^2$ and corresponding eigenvectors $\bm{r}^{(k)}$, which satisfy the orthogonality relation $\bm{r}^{(k')T} \gr \bm{r}^{(k)} = \delta_{k, k'}$. Here the $\bm{r}^{(k)}$ is equivalent to be defined as the solution to generalized eigen-problem:
\begin{align}
    \ci \bm{r}^{(k)} = \beta^2_k \gr \bm{r}^{(k)}. 
\end{align}
This is because $ \ci \bm{r}^{(k)} = \gr^{\frac{1}{2}} \nsr \gr^{\frac{1}{2}} \bm{r}^{(k)} = \beta_k^2 \gr^{\frac{1}{2}} \gr^{\frac{1}{2}} \bm{r}^{(k)} = \beta^2_k \gr \bm{r}^{(k)}$. The prefactor $\gr^{\frac{1}{2}}$ is introduced for later convenience. Eq.\,(\ref{eq:Reigen}) then allows us to define the related eigenproblem
\begin{align}
\left(\mathbf{I} + \frac{1}{\NS} \mathbf{R} \right)^{- 1}\!\!\! \gr^{\frac{1}{2}}\bm{r}^{(k)}  = \left( 1 + \frac{\beta_k^2}{\NS}  \right)^{-1} \gr^{\frac{1}{2}}\bm{r}^{(k)}
\label{eq:eigRinvSol}
\end{align}
Next, we note that $\mathbf{Q}$ is related to the matrix in brackets above via a \textit{generalized} similarity transformation defined by $\mathbf{B}$, Eq.\,(\ref{eq:Q}). In particular, $\mathbf{B}^T \mathbf{B} = \gr^{-\frac{1}{2}} \gr \gr^{-\frac{1}{2}} = \mathbf{I} \in \mathbb{R}^{K \times K}$, while we remark that $\mathbf{B}\mathbf{B}^T \neq \mathbf{I}$ since it is in $\mathbb{R}^{\infty \times \infty}$. This connection allow us to show that
\begin{align}
    \mathbf{Q} \mathbf{B} \gr^{\frac{1}{2}} \bm{r}^{(k)} = \mathbf{B} \left( \mathbf{I} + \frac{1}{\NS} \mathbf{R} \right)^{- 1}\!\!\! \mathbf{B}^T \mathbf{B} \gr^{\frac{1}{2}} \bm{r}^{(k)} = \frac{1}{1+\beta_k^2/S} \mathbf{B} \gr^{\frac{1}{2}} \bm{r}^{(k)}. 
    \label{eq:eiqQSol}
\end{align}
Comparing with Eq.\,(\ref{eq:eigQ}), we can now simply read off both the eigenvalues and eigenvectors of $\mathbf{Q}$,
\begin{align}
    \left. \begin{array}{rl}
        C_k & = \frac{1}{1+\beta_k^2/\NS}  \\
        \gh^{\frac{1}{2}}\mathbf{Y}^{(k)} & = \mathbf{B} \gr^{\frac{1}{2}} \bm{r}^{(k)} 
    \end{array} \right\}
    \implies \mathbf{Y}^{(k)} = \mathbf{T}^T \bm{r}^{(k)}
\end{align}
where we have used the definition of $\mathbf{B}$ from Eq.\,(\ref{eq:B}). The functions defined by the eigenvectors $\mathbf{Y}^{(k)}$ are automatically orthonormalized:
\begin{equation}
    \left\langle y^{(k_1)}, y^{(k_2)} \right\rangle_{p} = \left( \gh^{\frac{1}{2}} \mathbf{Y}^{(k_1)} \right)^T\!\! \left( \gh^{\frac{1}{2}} \mathbf{Y}^{(k_2)} \right) = \boldsymbol{r}^{(k_1)T} \gr^{\frac{1}{2}} \mathbf{B}^T \mathbf{B} \gr^{\frac{1}{2}} \boldsymbol{r}^{(k_2)} = \boldsymbol{r}^{(k_1)T} \gr \boldsymbol{r}^{(k_2)} = \delta_{k_1 k_2}. 
\end{equation}

\subsection{Noisy eigentasks from readout features}
\label{sec:noisy_ET}

We can now also discuss the interpretation of $\{\beta_k^2\}$ for a physical  system - in this case a quantum circuit - for which $\{\bm{r}^{(k)}\}$ are known. Consider a single run of the quantum system under finite shots $S$, which yields a single instance of the readout features $\bar{\bm{X}}(u)$. We can simply read off that an \textit{noisy} version of the $k$th eigentask, $\bar{y}^{(k)}(u)$ can be constructed as
\begin{align}
    \bar{y}^{(k)} (u) = \sum_{k' = 0}^{K-1} r_{k'}^{(k)} \bar{X}_{k'} (u)
\end{align}
which is equivalent to requiring the output weights $\bm{W} = \bm{r}^{(k)}$.The corresponding set of noisy function is also orthogonal, this is because $\ci \bm{r}^{(k)} = \beta^2_k \gr \bm{r}^{(k)}$ implies $\bm{r}^{(k)T} \ci \bm{r}^{(k')} = \beta^2_k \delta_{k, k'}$ and hence
\begin{align}
    \left\langle \bar{y}^{(k_1)}, \bar{y}^{(k_2)} \right\rangle_{p} = \boldsymbol{r}^{(k_1)T} \left( \gr +\frac{1}{S} \ci \right) \boldsymbol{r}^{(k_2)} = \left(1+\frac{\beta^2_k}{S}\right) \delta_{k_1 k_2}
\end{align}
This equation can be further decomposed into two parts. Let the linear transformation of noise $\bm{\xi}(u)$ by defining $\xi^{(k)}(u) = \frac{1}{\sqrt{S}} \sum_{k=0}^{K-1} r^{(k)}_{k'} \zeta_{k'}(u)$
\begin{align}
    \mathbb{E}_u [y^{(k_1)} y^{(k_2)} ]& = \left\langle y^{(k_1)}, y^{(k_2)} \right\rangle_{p} = \boldsymbol{r}^{(k_1)T} \gr \boldsymbol{r}^{(k_2)} = \delta_{k_1 k_2}, \\
    \mathbb{E}_u [\xi^{(k_1)} \xi^{(k_2)} ]& = \left\langle \xi^{(k_1)}, \xi^{(k_2)} \right\rangle_{p} = \frac{1}{S} \boldsymbol{r}^{(k_1)T} \ci \boldsymbol{r}^{(k_2)} = \frac{\beta^2_{k_1}}{S} \delta_{k_1 k_2}. \label{eq:ortnoise}
\end{align}
It means that the combination $\{\bm{r}^{(k)}\in\mathbb{R}^K\}_{k \in [K]}$ not only produces orthogonal eigentasks $\{ y^{(k)}(u) \}$ for signal, but also induces a set of orthogonal noise functions $\{ \xi^{(k)} (u) \}$.

If the quantum circuit can be run multiple times for a given $S$, multiple instances of $\bar{\bm{X}}(u)$ can be obtained, from each of which an estimate of the $k$th eigentask $\bar{y}^{(k)}(u)$ can be constructed. The expectation value of these estimates then simply yields
\begin{align}
    \mathbb{E}[\bar{y}^{(k)} (u)] = \sum_{k' = 0}^{K-1} r_{k'}^{(k)} \mathbb{E}[\bar{X}_{k'} (u)] = \sum_{k' = 0}^{K-1} r_{k'}^{(k)} {x}_{k'} (u) = {y}^{(k)}(u)
\end{align}

If we have access to only a single instance of $\bar{\bm{X}}(u)$, however, and thus only one estimate $\bar{y}^{(k)}(u)$ (as $y^{(k)}(u)$ and $\bar{y}^{(k)}(u)$ depicted in Fig.\,\ref{fig:app_Features_and_Capacity}), it is useful to know the expected error in this estimate. This error can be extracted from Eq.\,(\ref{eq:Cf}). In particular, requiring $\mathbf{Y}^{(k)} = \mathbf{T}^T \bm{r}^{(k)}$, we have
\begin{align}
    &\frac{ \left\| \gh^{\frac{1}{2}} \mathbf{T}^T \bm{r}^{(k)} - \gh^{\frac{1}{2}} \mathbf{Y}^{(k)} \right\|^2 + \frac{1}{\NS} \bm{r}^{(k)T} \ci \bm{r}^{(k)} }{\mathbf{Y}^{(k)T} \gh \mathbf{Y}^{(k)}} = \frac{1}{S} \bm{r}^{(k)T} \ci \bm{r}^{(k)} = \frac{\beta_k^2}{S}. 
\end{align}
This mean squared error in using $\bar{y}^{(k)}(u)$ to estimate ${y}^{(k)}(u)$ over the domain of $u$ decreases to zero for $S\to\infty$ as expected, since the noise in $\bar{\bm{X}}$ decreases with $S$. However, $\beta_k^2$ defines the $S$-independent contribution to the error. In particular, this indicates that at a given $S$, certain functions with lowers NSR eigenvalues $\beta_k^2$ may be better approximated using this physical system than others. We present in Fig.\,\ref{fig:app_Features_and_Capacity} the measured features $\bar{\bm{X}}$, the eigentasks $\bm{y}$ and their $S$-finite version $\bar{\bm{y}}$ in a 6-qubit Hamiltonian based system. The associated eigen-NSR spectrum, expressive capacity, and total correlations are also depicted for both CS $J \neq 0$ and PS $J = 0$.

\subsection{Expressive capacity}
\label{app:EC}
Given an arbitrary set of complete orthonormal basis functions $f_{\ell} (u) = \sum_{j = 0}^{\infty} (\mathbf{Y}_{\ell})_j u^j$,
\begin{equation}
     \langle f_{\ell}, f_{\ell'} \rangle_p = \left( \gh^{\frac{1}{2}} \mathbf{Y}_{\ell} \right)^T \left( \gh^{\frac{1}{2}} \mathbf{Y}_{\ell'} \right) = \delta_{\ell \ell'} .
\end{equation}
The total capacity is independent of the basis choice
\begin{align}
    \EC(\NS) & = \sum_{\ell = 0}^{\infty} C [f_{\ell}] = \sum_{\ell = 0}^{\infty} \mathbf{Y}_{\ell}^T \gh^{\frac{1}{2}} \left( \gh^{\frac{1}{2}} \mathbf{T}^T \left( \mathbf{T}\gh\mathbf{T}^T + \frac{1}{\NS} \ci \right)^{- 1} \!\!\! \mathbf{T}\gh^{\frac{1}{2}} \right) \gh^{\frac{1}{2}} \mathbf{Y}_{\ell} \nonumber\\
    & = \mathrm{Tr} \left( \gh^{\frac{1}{2}} \mathbf{T}^T \left( \mathbf{T}\gh\mathbf{T}^T + \frac{1}{\NS} \ci \right)^{- 1} \!\!\! \mathbf{T}\gh^{\frac{1}{2}} \right) = \mathrm{Tr} \left( \left( \gr + \frac{1}{\NS} \ci \right)^{- 1} \!\!\! \gr \right) = \sum_{k=0}^{K-1} \frac{1}{1 + \frac{\beta_k^2}{\NS}}. 
\end{align}

\subsection{Estimation in case of nonlinear functions after linear output layer}
\label{app:ComplxNonLin}

Usually, instead of taking the linear transformation $\bm{W} \cdot \bar{\bm{X}}$, the training process can involve some complicated nonlinear activation functions or classical kernel, which may also be fed into a non-quadratic nonlinear loss function afterwards. These two processes can be unified to be $\sigma_{\mathrm{NL}}(\bar{\bm{X}}(u))$ with any smooth function $\sigma_{\rm{NL}}$. In this subsection, we show how to translate our result obtaining from quadratic nonlinear function Eq.\,(\ref{eq:defIPC}) into a more general loss function with form of
\begin{align}
    \mathscr{L} = \mathbb{E}_u[\sigma_{\mathrm{NL}}(\bar{\bm{X}})]
\end{align} 
Now let us first transform all noisy measured features $\{\bar{X}_k\}$ into the naturally orthogonal basis of signal $\{y^{(k)}\}$ and noise $\{\xi^{(k)}\}$. 
\begin{align}
    \bar{X}_{k'}(u) \equiv \sum_{k=0}^{K-1} \Gamma_{k'k} (y^{(k)}(u) + \xi^{(k)} (u)) ,
\end{align}
such transformation of $\bm{\Gamma} \in \mathbb{R}^{K \times K}$ must uniquely exist, this is because all $K$ of $\{\bm{r}^{(k)}\}$ are linearly independent. Recall Eq.\,(\ref{eq:ortnoise}) claims that $\mathbb{E}_u [\xi^{(k)}] = 0$ and $\mathbb{E}_u [\xi^{(k)} \xi^{(k')} ]=\beta^2_{k} \delta_{k k'}/S$, we can deal with the nonlinearity by taking the cumulant expansion up to the quadratic term, and we get
\begin{align}
    \mathscr{L} & = \mathbb{E}_u [\sigma_{\mathrm{NL}} (\bar{\bm{X}})] = \mathbb{E}_u [\sigma_{\mathrm{NL}} (\mathbf{\Gamma} \bar{\bm{y}})] = \mathbb{E}_u \! \left[ \sigma_{\mathrm{NL}} \! \left( \sum_k \Gamma_{0, k} (y^{(k)} + \xi^{(k)}), \cdots, \sum_k \Gamma_{K - 1, k} (y^{(k)} + \xi^{(k)}) \right) \right] \nonumber\\
    & = \mathbb{E}_u [\sigma_{\mathrm{NL}} (\mathbf{\Gamma} \bm{y})] + \sum_{k = 0}^{K - 1} \mathbb{E}_u \left[ \frac{\partial \sigma_{\mathrm{NL}}}{\partial y^{(k)}} \xi^{(k)} \right] + \frac{1}{2} \sum_{k_1 = 0}^{K - 1} \sum_{k_2 = 0}^{K - 1} \mathbb{E}_u \left[ \frac{\partial^2 \sigma_{\mathrm{NL}}}{\partial y^{(k_1)} \partial y^{(k_2)}} \xi^{(k_1)} \xi^{(k_2)} \right] + O\!\left(\frac{1}{S^2}\right) \nonumber\\
    & \approx \mathbb{E}_u [\sigma_{\mathrm{NL}} (\mathbf{\Gamma} \bm{y})] + \frac{1}{2} \sum_{k_1 = 0}^{K - 1} \sum_{k_2 = 0}^{K - 1} \mathbb{E}_u \left[ \frac{\partial^2 \sigma_{\mathrm{NL}}}{\partial y^{(k_1)} \partial y^{(k_2)}} \bm{r}^{(k_1)T} \mathbf{\Sigma} \bm{r}^{(k_2)} \right], \label{eq:Lapprox}
\end{align}
where the first order terms vanish due to Hoeffding inequality again. We then make a further approximation of Eq.\,(\ref{eq:Lapprox}) by replacing the $\xi^{(k_1)} \xi^{(k_2)}$ with its $u$-average $\mathbb{E}_u [\xi^{(k_1)} \xi^{(k_2)}] = \delta_{k_1 k_2} \beta_{k_1}^2 / S$:
\begin{equation}
   \mathscr{L} \approx \mathbb{E}_u [\sigma_{\mathrm{NL}} (\mathbf{\Gamma} \bm{y})] + \sum_{k = 0}^{K - 1} \frac{\beta_k^2}{S} \cdot \mathbb{E}_u \! \left[\frac{\partial^2 \sigma_{\mathrm{NL}}}{(\partial y^{(k)})^2} \right]. \label{eq:Lapprox-1}
\end{equation} 
In fact, any of the second terms can be further simplified by chain rule: $\mathscr{L} \approx \mathbb{E}_u [\sigma_{\mathrm{NL}} (\mathbf{\Gamma} \bm{y})] + \sum_{k} \frac{\beta_k^2}{S} \cdot \mathbb{E}_u [(\mathbf{\Gamma}^T \nabla_{\bm{x}}^2 \sigma_{\mathrm{NL}} \mathbf{\Gamma})_{k k}]$. The approximation in  Eq.\,(\ref{eq:Lapprox-1}) is rough, but it still gives us a sufficient reason to do the following manipulation: for optimized $\mathscr{L}$, the dependence on $y^{(k)}$ with $\beta_k^2 / \NS>1$ will be strongly suppressed in large-$N$ limit, hence we can pre-exclude the eigentasks whose $\beta_k^2 / \NS>1$.

Let us use one typical example, the widely used logistic regression in classification, to illustrate our argument here. As what we will introduce in Appendix \ref{sec:Single_step_QRC}, the target function is the conditional probability distribution $f(u) := \Pr[u \in C_1|u]$ in such classification model (see Eq.\,(\ref{eq:classifier_approx})), and then there is one more layer of softmax and cross-entropy function acting on linear map $\mathscr{L} = \mathbb{E}_u[\operatorname{H} (f(u), \sigma(\bm{W} \cdot \bar{\bm{X}}(u)))]$ where $\sigma$ is sigmoid function (\textit{e.g.} softmax function $\sigma(z) = 1/(1+\mathrm{exp}(-z))$), and $\operatorname{H}(p,q) = - p \ln q - (1-p) \ln (1-q)$ is the cross-entropy. Especially, any linear combination of $\{\bar{X}_k\}$ can be translated into linear combination
\begin{align}
    \bm{W} \cdot \bar{\bm{X}}(u) \equiv \sum_{k=0}^{K-1} \Omega_{k} \cdot (y^{(k)}(u) + \xi^{(k)} (u)) ,
\end{align}
Again, such vector $\bm{\Omega} = \bm{\Gamma}^T \bm{W}$ must also uniquely exist. For any $\sigma_{\mathrm{NL}} = g(\bm{W} \cdot \bm{x})$, one always have $\mathbf{\Gamma}^T \nabla_{\bm{x}}^2 \sigma_{\mathrm{NL}} \mathbf{\Gamma} = g''(\bm{\Omega}\cdot \bm{y}) \mathbf{\Omega}^T \mathbf{\Omega}$: 
\begin{align}
    \mathscr{L} & \approx 
    \mathbb{E}_u \!\left[\operatorname{H} \! \left(f, \sigma\!\left(\bm{\Omega} \cdot \bm{y}\right)\right)\right] + \left(\sum_{k=0}^{K-1} \frac{\beta_k^2}{\NS} \Omega^2_{k}\right) \cdot \mathbb{E}_u \!\left[ \sigma(\bm{\Omega} \cdot \bm{y}) (1-\sigma(\bm{\Omega} \cdot \bm{y})) \right]. \label{eq:LapproxCls}
\end{align}
It helps us read from the prefactor $\beta_k^2 / \NS$ induces a natural regularization on $\Omega_{k}$ in loss function, in addition to the $S$-infinity term $\lim_{S \to \infty} \mathscr{L} = \mathbb{E}_u \!\left[\operatorname{H} \! \left(f, \sigma\!\left(\bm{\Omega} \cdot \bm{y}\right)\right)\right] $. We will leave the detailed discussion of this important application in Appendix \ref{app:PCA} and Appendix \ref{sec:Single_step_QRC}.

\subsection{Proof that the Gram matrix $\gr$ is full rank} 
\label{app:Function-independence}
Recall that before we analytically find the eigenvectors of $\mathbf{Q}$, we first show that the matrix $\gr$ is invertible. 
It comes from that all $K$ readout features $\{x_k(u)\}_{k \in [K]}$ being linear independent is entirely equivalent to the full-rankness of the corresponding Gram matrix $\mathrm{Rank}(\gr)=K$. Thanks to the linearity of readout, we can show such linear independence by contradiction. Suppose on the contrary there exists coefficients $\{ c_k \}_{k \in [K]}$ such that
\begin{align}
    \sum_{k = 0}^{K-1} c_k x_k (u) = \mathrm{Tr} \left\{ \left( \sum_{k = 0}^{K-1} c_k \hat{E}_k  \right) \mathcal{U} (u) \hat{\rho}_0 \right\} = 0. 
\end{align}
However, this means that the quantum observable $\sum_{k = 0}^{K-1} c_k \hat{E}_k$ is a zero-expectation readout-qubit quantity for any state $\mathcal{U} (u) \hat{\rho}_0$ under arbitrary input $u$, which is impossible. This shows the linear independence. Furthermore, we then argue that it ensures $\gr$ has no non-trivial null space. This is because that any $\{ c_k \}_{k \in [K]}$ will satisfy
\begin{align}
    \sum_{k_1,k_2=1}^{K} c_{k_1} c_{k_2} (\gr)_{k_1, k_2} =  \int \left( \sum_{k_1=1}^{K} c_{k_1} x_{k_1} (u) \right)\!\! \left( \sum_{k_2=1}^{K} c_{k_2} x_{k_2} (u) \right) p(u) \dd u = \left\langle \sum_{k=0}^{K-1} c_{k} x_{k}, \sum_{k=0}^{K-1} c_{k} x_{k} \right\rangle_p. 
\end{align}
where the RHS is the norm of function $\sum_{k=0}^{K-1} c_{k} x_{k} (u)$. The summation $\sum_{k_1,k_2=1}^{K} c_{k_1} c_{k_2} (\gr)_{k_1, k_2}=0$ vanishes if and only if function $\sum_{k=0}^{K-1} c_{k} x_{k} (u)$ is a zero function. That is why the linear independence of features $\{ c_k \}_{k \in [K]}$ is equivalent to that symmetric matrix $\gr$ has no zero eigenvalues, namely $\operatorname{Rank}(\gr)=K$. 

\subsection{Simplifying the noise-to-signal matrix and its eigenproblem}

We have shown that the problem of obtaining the eigentasks for a generic quantum system, and deducing its expressive capacity under finite measurement resources, can be reduced simply to solving the eigenproblem of its noise-to-signal matrix $\mathbf{R}$, Eq.\,(\ref{eq:Reigen}). Note that constructing $\mathbf{R}=\gr^{-\frac{1}{2}}\mathbf{V}\gr^{-\frac{1}{2}}$ requires computing the inverse of $\gr$. However, $\gr$ can have small (although always nonzero) eigenvalues, especially for larger systems, rendering it ill-conditioned and making the computation of $\mathbf{R}$ numerically unstable. Fortunately, certain simplifications can be made to derive an equivalent eigenproblem that is much easier to solve. Practically, the probability representation is native to measurement schemes in contemporary quantum processors, and therefore minimizes the required post-processing of readout features obtained from a real device. More importantly, the strength of the probability representation lies in the fact that it renders the second-order moment matrix $\mathbf{D}$ diagonal. In particular,
\begin{align}
    (\mathbf{D})_{k_1k_2} = \left\{
    \begin{array}{cc}
        \sum_{k = 0}^{K-1} (\gr)_{k k_1}, &\text{if } k_1=k_2 \\
        0, & \text{if } k_1 \neq k_2
    \end{array} \right.~~~
    \text{(in~probability~representation~of~readout~features)}
\end{align}

Using $\mathbf{V} =\mathbf{D}-\gr$, we can rewrite the eigenproblem for $\mathbf{R}$,
\begin{align}
    \nsr \left( \gr^{\frac{1}{2}}\bm{r}^{(k)} \right) & = \beta_k^2 \gr^{\frac{1}{2}}\bm{r}^{(k)} \nonumber \\
    \implies \gr^{-\frac{1}{2}}(\mathbf{D}-\gr)\gr^{-\frac{1}{2}} \left( \gr^{\frac{1}{2}}\bm{r}^{(k)} \right) &= \beta_k^2 \gr^{\frac{1}{2}}\bm{r}^{(k)} \nonumber \\
    \implies \gr^{-1}\mathbf{D} \bm{r}^{(k)} &= (1+\beta_k^2) \bm{r}^{(k)}
\end{align}
Finally, considering the inverse of the matrix on the left hand side, we obtain the simplified eigenproblem for the matrix $\mathbf{D}^{-1}\gr$,
\begin{align}
    \mathbf{D}^{-1}\gr \bm{r}^{(k)} = (1+\beta_k^2)^{-1} \bm{r}^{(k)} \equiv \alpha_k \bm{r}^{(k)},
    \label{eq:DinvGeigen}
\end{align}
which shares eigenvectors with $\nsr$, and whose eigenvalues are a simple transformation of the NSR eigenvalues $\beta_k^2$. Importantly, constructing $\mathbf{D}^{-1}\gr$ no longer requires calculating any powers of $\gr$, and when further choosing readout features in the probability representation, it relies only on the inversion of a simple diagonal matrix $\mathbf{D}$.

The matrix $\mathbf{D}^{-1}\gr$ has significance in spectral graph theory, when interpreting the Gram matrix $\gr$ as the adjacency matrix of a weighted graph. This connection is elaborated upon in Appendix \ref{sec:graph}.

\subsection{Connections to spectral graph theory}
\label{sec:graph}

Let us have a small digression to the graphic theoretic meaning of $\gr$ and $\mathbf{\mathbf{D}}^{-1} \gr$. Now we consider a weighted graph with adjacency matrix $\gr$. In spectral graph theory, the matrix $\mathbf{\mathbf{D}}^{- 1} \gr$ is exactly the random walk matrix associated with graph $\gr$, and then the second order matrix $\mathbf{D}$ happens to be the \textit{degree matrix} of this graph since $(\mathbf{D})_{k k}= \sum_{k' = 0}^{K-1} (\mathbf{G})_{k k'}$. Then the eigentask combination coefficient $\boldsymbol{r}^{(k)}$ is precisely the right eigenvector of random walk matrix.  Another concept associated with a graph is $\mathbf{I} - \mathbf{D}^{-\frac{1}{2}} \gr \mathbf{D}^{-\frac{1}{2}}$, the \textit{normalized Laplacian matrix} of $\gr$, while the matrix $\mathbf{D}^{-\frac{1}{2}} \gr \mathbf{D}^{-\frac{1}{2}}$ is always referred to be \textit{normalized adjacency matrix} in many literatures. The eigenproblem of normalized adjacency matrix can also be solved easily, because 
\begin{align}
    \mathbf{D}^{-\frac{1}{2}} \gr \mathbf{D}^{-\frac{1}{2}} \left( \mathbf{D}^{\frac{1}{2}} \bm{r}^{(k)} \right) = \mathbf{D}^{\frac{1}{2}} \mathbf{D}^{-1} \gr \bm{r}^{(k)} = \alpha_k \left( \mathbf{D}^{\frac{1}{2}} \bm{r}^{(k)} \right).
\end{align}
From perspective of spectral graph theory, roughly speaking, a reservoir with stronger ability to resist noise are those who has more ``bottlenecks" in graph $\gr$'s connectivity. The extreme case is supposing that $\alpha_k=1$ (or $1-\alpha_k=0$) for all $k$. According the basic conclusion in spectral graph theory, the normalized Laplacian matrix has $K$ zero eigenvalues iff the graph $\gr$ is fully disconnected. This gives us the condition when noisy information capacity obtain its upper bound $K$: there exists a partition $\{ \mathrm{Dom}_k \}_{k \in [K]}$ of domain $\mathrm{Dom}=[-1,1]$ such that $\hat{\rho}_{kk}(u) = 1$ iff $u \in \mathrm{Dom}_k$.


\section{Spectral analysis based on finite statistics}
\label{app:Spectral_finite_statistics}

While Eq.\,(\ref{eq:DinvGeigen}) is a numerically simpler eigenproblem to solve than Eq.\,(\ref{eq:Reigen}), it still requires the approximation of $\gr$ (recall that $\mathbf{D}$ can be obtained from $\gr$) from readout features $\bar{\bm{X}}(u)$ under finite sampling, due to the finiteness of shots $\NS$, the number of input points $N$, and also the number of realizations of readout features for a given $\NS$. 
\rev{To be more precise, in experiment one only has access to measured features sampled at finite-$S$ $\bar{\bm{X}}$ (indeed, this distinction is the underlying premise of this article).  However, in Eq.\,\eqref{eq:eigenprob} $\gr$ and $\ci$ are defined with respect to the ideal $\bm{x}$. Let $\widetilde{\gr} \equiv \mathbb{E}_{\UI} [\bar{\bm{X}}\bar{\bm{X}}^T]$ and $\widetilde{\ci} \equiv \mathbb{E}_{\UI} [\mathrm{diag}(\bar{\bm{X}})-\bar{\bm{X}}\bar{\bm{X}}^T]$. The objective of Appendix \ref{app:tilde_correction} is showing that the eigen-analysis $\{\beta^2_k, \bm{r}^{(k)}\}$ can be accurately substituted with
\begin{align}
    \beta_k^2 = \frac{S \cdot \tilde{\beta}_k^2}{(S-1)-\tilde{\beta}_k^2}, \label{eq:btilde}
\end{align}
and $\bm{r}^{(k)} = \tilde{\bm{r}}^{(k)}$ from solving generalized eigenvalue problem $\widetilde{\ci} \tilde{\bm{r}}^{(k)} = \tilde{\beta}_k^2 \widetilde{\gr} \tilde{\bm{r}}^{(k)}$. }
In what follows, we show how an approximation $\widetilde{\gr}_N$ of $\gr$ can be constructed from finitely-sampled readout features, as relevant for practical quantum devices. Secondly, we also describe an approach to obtain the eigentasks $y^{(k)}(u)$ and corresponding NSR eigenvalues $\beta_k^2$ that avoids explicit construction of the Gram matrix, and is thus even more numerically robust.

\subsection{Approximating eigentasks and NSR eigenvalues under finite $\NS$ and $N$}
\label{app:tilde_correction}

For practical computations, readout features $\bar{\bm{X}}(u)$ from the quantum system for finite $\NS$ can be computed for a discrete set of $u^{(n)} \in [-1,1]$ for $n = 1,\ldots,N$. Labelling the corresponding readout features $\bar{\bm{X}}(u^{(n)})$, we can define the \textit{regression matrix} constructed from these readout features,
\begin{equation}
    \regmat \equiv (\bar{\bm{X}} (u^{(1)}), \bar{\bm{X}} (u^{(2)}), \cdots, \bar{\bm{X}} (u^{(N)}))^T = \left(\begin{array}{ccc}
        \bar{X}_0 (u^{(1)}) & \cdots & \bar{X}_{K-1} (u^{(1)})\\
        \vdots &  & \vdots\\
        \bar{X}_0 (u^{(N)}) & \cdots & \bar{X}_{K-1} (u^{(N)})
    \end{array}\right) .
\end{equation}
Here, $\regmat \in \mathbb{R}^{N\times K}$, with subscript $N$ indicating its construction from a finite set of $N$ inputs, is a random matrix due to the stochasticity of readout features; in particular it can be written as:
\begin{align}
    \regmat = \RM  + \frac{1}{\sqrt{\NS}}\mathbf{Z}(\RM)
\end{align}
where $(\RM)_{nk} = \mathbb{E}[\bar{X}_k(u^{(n)})] = x_k(u^{(n)})$, and $\mathbf{Z}$ is the centered multinomial stochastic process, so that $\mathbb{E}[\widetilde{\mathbf{F}}_N] = \RM$.

Using this regression matrix $\regmat$, we can obtain an estimation of the Gram matrix and second order moment matrix, which we denote $\widetilde{\gr}_N$ and $\widetilde{\mathbf{D}}_N$, and whose matrix elements are defined via
\begin{align}
     (\widetilde{\gr}_N)_{k_1 k_2} & \equiv \frac{1}{N}\sum_{n=1}^N \bar{X}_{k_1}(u^{(n)})\bar{X}_{k_2}(u^{(n)}) = \frac{1}{N} (\regmat^T \regmat)_{k_1k_2} \approx \int \bar{X}_{k_1}(u)\bar{X}_{k_2}(u) p(u) \dd u, \\
     (\widetilde{\mathbf{D}}_N)_{k_1 k_2} & \equiv \delta_{k_1, k_2} \frac{1}{N}\sum_{n=1}^N \bar{X}_{k_1}(u^{(n)}) \approx \delta_{k_1, k_2} \int \bar{X}_{k_1}(u) p(u) \dd u.
\end{align}
While the quantities $\widetilde{\gr}_N$ and $\widetilde{\mathbf{D}}_N$ are computed from stochastic readout features, their stochastic contributions are suppressed in the large $N$ limit by the Hoeffding inequality for sums of bounded stochastic variables. In particular, we can define their deterministic limit for $N \to \infty$, according to Eq.\,(\ref{eq:Lambdatilde}), as
\begin{align}
    \widetilde{\gr} &\equiv \lim_{N \to \infty} \frac{1}{N} (\regmat^T \regmat)_{k_1k_2} = \gr + \frac{1}{\NS} \ci = \gr + \frac{1}{\NS} (\mathbf{D}-\gr),
    \\ \widetilde{\mathbf{D}} &\equiv \lim_{N \to \infty} \widetilde{\mathbf{D}}_N = \mathbf{D}. 
\end{align}
Inverting the above expressions allow us to express the Gram matrix $\gr$ and second-order moment matrix $\mathbf{D}$ in terms of the estimates $\widetilde{\mathbf{G}}$ and $\widetilde{\mathbf{D}}$ computed using a finite number of shots $\NS$,
\begin{align}
    \gr &= \frac{\NS}{\NS-1}\widetilde{\mathbf{G}} - \frac{1}{\NS-1}\widetilde{\mathbf{D}},  \\
    \mathbf{D} &= \widetilde{\mathbf{D}}.
\end{align}
We see that to lowest order in $\frac{1}{\NS}$, $\gr \approx \widetilde{\gr}$ and $\mathbf{D} \approx \widetilde{\mathbf{D}}$, which is what one might expect naively. However, we clearly see that the estimation of $\gr$ can be improved by including a higher-order correction in $\frac{1}{\NS}$. This contribution arises due to the highly-correlated nature of noise and signal for quantum systems: we are able to estimate the noise matrix $\widetilde{\mathbf{G}}$ and $\widetilde{\mathbf{D}}$ using knowledge of the readout features, and correct for the contribution to $\widetilde{\gr}$ and $\widetilde{\mathbf{D}}$ that arises from this noise matrix. We will see that this contribution will be important in more accurately approximating quantities of interest derived from $\gr$, $\mathbf{D}$.

\begin{figure}[t]
    \centering
    \includegraphics[width=0.5\textwidth]{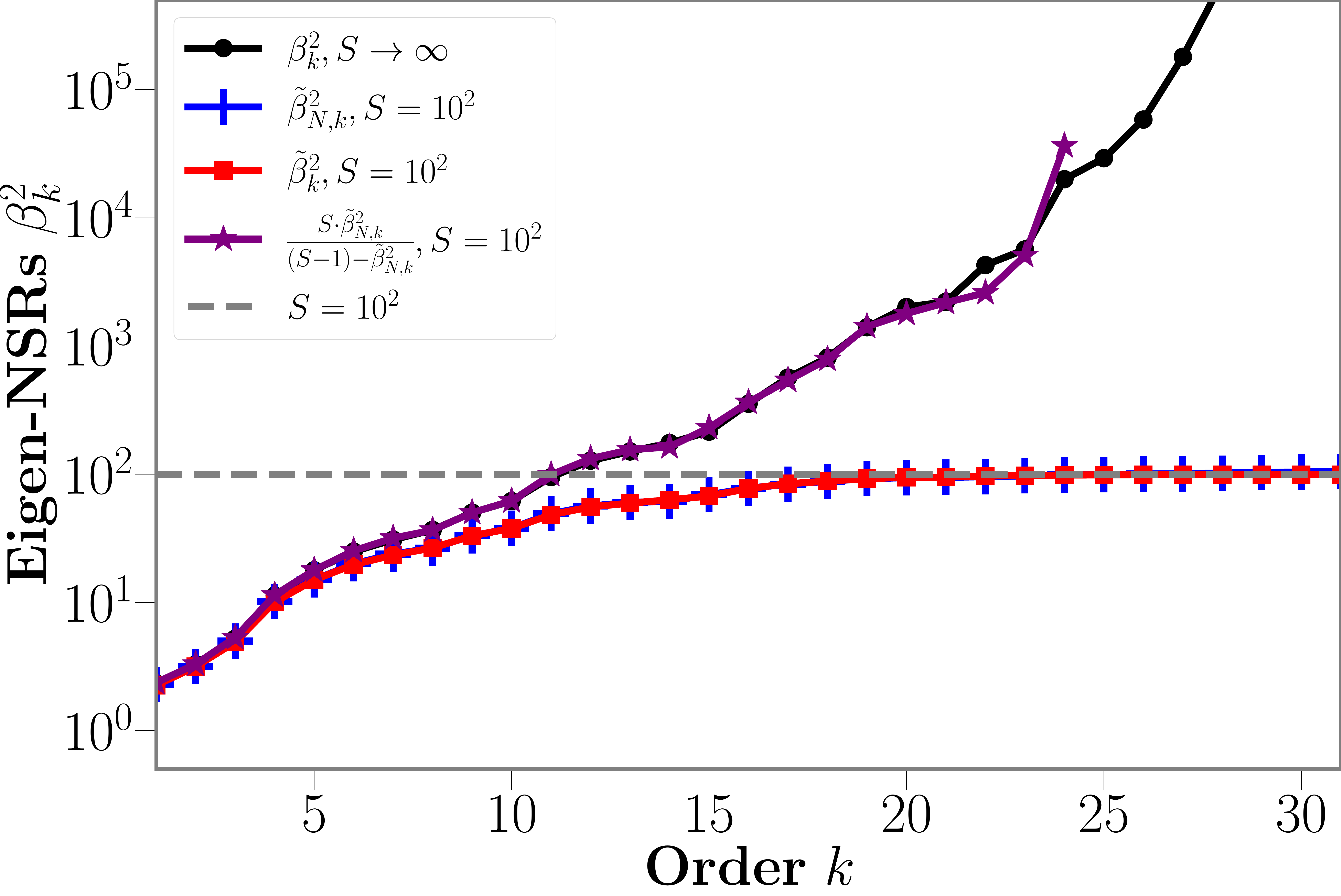}
    \caption{Eigen-analysis in $L=5$ H-ansatz system by taking $\NS=10^2$ shots on each of $N = 10^4$ samples, with true eigen-noise-to-signal ratios $\beta_k^2$ (black), $\NS$-finite sampled $\tilde{\beta}_{N,k}^2$ (blue) and corrected $(\NS \cdot \tilde{\beta}_{N,k}^2)/((\NS - 1) - \tilde{\beta}_{N,k}^2)$ (purple). $\tilde{\beta}_k^2$, the large $N$ limit of $\tilde{\beta}_{N,k}^2$ is also plotted in red for comparison. The data correction is necessary since all $\tilde{\beta}_{N,k}^2$ are below the $S=10^2$, and the corrected data show much better performance even if $\beta_k^2 \gg \NS$. The estimated line (in purple) are cutoff at $k=25$ since all sampled $\tilde{\beta}_{N,k}^2$ after that are larger the $S-1$ so that they are not correctable. 
    }
    \label{fig:NSR_Tilde_Renormalization}
\end{figure}

To this end, we recall that our ultimate aim is not just to estimate $\gr$ and $\mathbf{D}$, but to solve the eigenproblem of Eq.\,(\ref{eq:DinvGeigen}). Using the above relation, we can then establish $\widetilde{\mathbf{D}}^{-1}\widetilde{\mathbf{G}} = \frac{\NS-1}{\NS}\mathbf{D}^{-1}\gr +\frac{1}{\NS} \mathbf{I}$, and write Eq.\,(\ref{eq:DinvGeigen}) in a form entirely in terms of $\widetilde{\mathbf{G}}$ and $\widetilde{\mathbf{D}}$,
\begin{align}
    \mathbf{D}^{-1}\gr \bm{r}^{(k)} &= (1+\beta_k^2)^{-1} \bm{r}^{(k)}, \nonumber \\
    \implies \widetilde{\mathbf{D}}^{-1}\widetilde{\mathbf{G}}\bm{r}^{(k)} &= \left[\frac{\NS-1}{\NS}(1+\beta_k^2)^{-1} +\frac{1}{\NS} \right]\bm{r}^{(k)}. 
\end{align}
Note that the final form is conveniently another eigenproblem, now for the finite-$\NS$ matrix $\widetilde{\mathbf{D}}^{-1}\widetilde{\mathbf{G}}$:
\begin{align}
    \widetilde{\mathbf{D}}^{-1}\widetilde{\mathbf{G}} \tilde{\bm{r}}^{(k)} = (1+\tilde{\beta}_k^2)^{-1} \tilde{\bm{r}}^{(k)} \equiv \tilde{\alpha}_k \tilde{\bm{r}}^{(k)},
    \label{eq:DinvGeigenFinS}
\end{align}
whose eigenvalues and eigenvectors can be easily related to the desired eigenvalues $\beta_k^2$ and eigenvectors $\bm{r}^{(k)}$ of Eq.\,(\ref{eq:DinvGeigen}). Following some algebra, we find:
\begin{align}
    \beta_k^2 &= \frac{S}{(\NS-1) -\tilde{\beta}_k^2}\cdot \tilde{\beta}_k^2 = \tilde{\beta}_k^2 +  
    \sum_{j=1}^{\infty} \tilde{\beta}_k^2 \left( 1+\tilde{\beta}_k^2 \right)^{j} \left( \frac{1}{S} \right)^j, \label{eq:betaEst} \\
    \bm{r}^{(k)} &= \tilde{\bm{r}}^{(k)}. 
    \label{eq:rEst}
\end{align}
From Eq.\,(\ref{eq:betaEst}), we see that to lowest order in $\frac{1}{\NS}$, $\beta_k^2 \approx \tilde{\beta}_k^2$. However, this expression also supplies corrections to higher orders in $\frac{1}{S}$, which are non-negligible even for $\beta_k^2 < S$, as we see in example of Fig.\,\ref{fig:NSR_Tilde_Renormalization}. In contrast, the estimated eigenvectors $\tilde{\bm{r}}^{(k)}$ to \textit{any} order in $\frac{1}{S}$ equal the desired eigenvectors ${\bm{r}}^{(k)}$ without any corrections. 

Of course, in practice we do not have access to the matrices $\widetilde{\gr}$ and $\widetilde{\mathbf{D}}$, as these are only defined precisely in the limit where $N\to\infty$. However, for large enough $N$, we can approximate these matrices to lowest order by their finite $N$ values, $\widetilde{\gr} = \widetilde{\gr}_N + \mathcal{O}\left(\frac{1}{N}\right)$ and $\widetilde{\mathbf{D}} = \widetilde{\mathbf{D}}_N + \mathcal{O}\left(\frac{1}{N}\right)$. 
Then, the eigenproblem in Eq.\,(\ref{eq:DinvGeigenFinS}) can be expressed in the final form,
\begin{align}
    \widetilde{\mathbf{D}}_N^{-1} \widetilde{\gr}_N \tilde{\bm{r}}_{N}^{(k)} = (1 + \tilde{\beta}_{N, k}^2)^{-1} \tilde{\bm{r}}_{N}^{(k)} \equiv \tilde{\alpha}_{N,k} \tilde{\bm{r}}_{N}^{(k)}, 
    \label{eq:DinvGeigenApprox}
\end{align}
where the eigenvalues $\tilde{\beta}_{N,k}^2, \tilde{\alpha}_{N,k}$ and eigenvectors $\tilde{\bm{r}}_{N}^{(k)}$ in the large $N$ limit must satisfy

\begin{align}
    \lim_{N \to \infty} \tilde{\beta}_{N,k}^2 = \tilde{\beta}_k^2, \quad \lim_{N \to \infty} \tilde{\alpha}_{N,k} = \tilde{\alpha}_k, \quad 
    \lim_{N \to \infty} \tilde{\bm{r}}_{N}^{(k)} = \tilde{\bm{r}}^{(k)} \equiv \bm{r}^{(k)}. 
\end{align}
Here the invertibility of the empirically-computed matrix $\widetilde{\mathbf{D}}_N$ required for Eq.\,(\ref{eq:DinvGeigenApprox}) is numerically checked, based on which we can establish a better numerical method in Appendix \ref{sec:GramMatrix-free}. 

Eq.\,(\ref{eq:DinvGeigenApprox}) represents the eigenproblem whose eigenvalues $\tilde{\beta}_{N,k}^2$ and eigenvectors $\tilde{\bm{r}}_{N}^{(k)}$ we actually calculate. For large enough $N$ and under finite $\NS$, we can use these as valid approximations to the eigenvalues and eigenvectors of Eq.\,(\ref{eq:DinvGeigenFinS}). This finally enables us to directly estimate the $N,\NS \to \infty$ quantities $\beta_k^2$ and $\bm{r}^{(k)}$ using Eqs.\,(\ref{eq:betaEst}), (\ref{eq:rEst}):
\begin{align}
    \beta_k^2 &\approx \frac{\NS \cdot \tilde{\beta}_{N,k}^2}{(\NS - 1) - \tilde{\beta}_{N,k}^2} = \frac{1 - \tilde{\alpha}_{N,k}}{\tilde{\alpha}_{N,k} - \frac{1}{\NS}}, \\
    \bm{r}^{(k)} &\approx \tilde{\bm{r}}_N^{(k)}.  
\end{align}
It is clear that the approximation of $\beta_k^2$ to lowest order will be an underestimate, as the contribution of order $\frac{1}{\NS}$ is positive. In Fig.\,\ref{fig:Tilde_Renormalization}, we plot the estimated eigenvectors $\tilde{\bm{r}}_N^{(k)}$ computed under finite statistics ($N=300,S=1000$, where these two numbers are relevant for IBM quantum processors) in H-encoding, together with the $N,\NS \to \infty$ eigenvectors ${\bm{r}}^{(k)}$, and the estimated eigenvalues.

\begin{figure}
    \centering
    \includegraphics[width=1.0\textwidth]{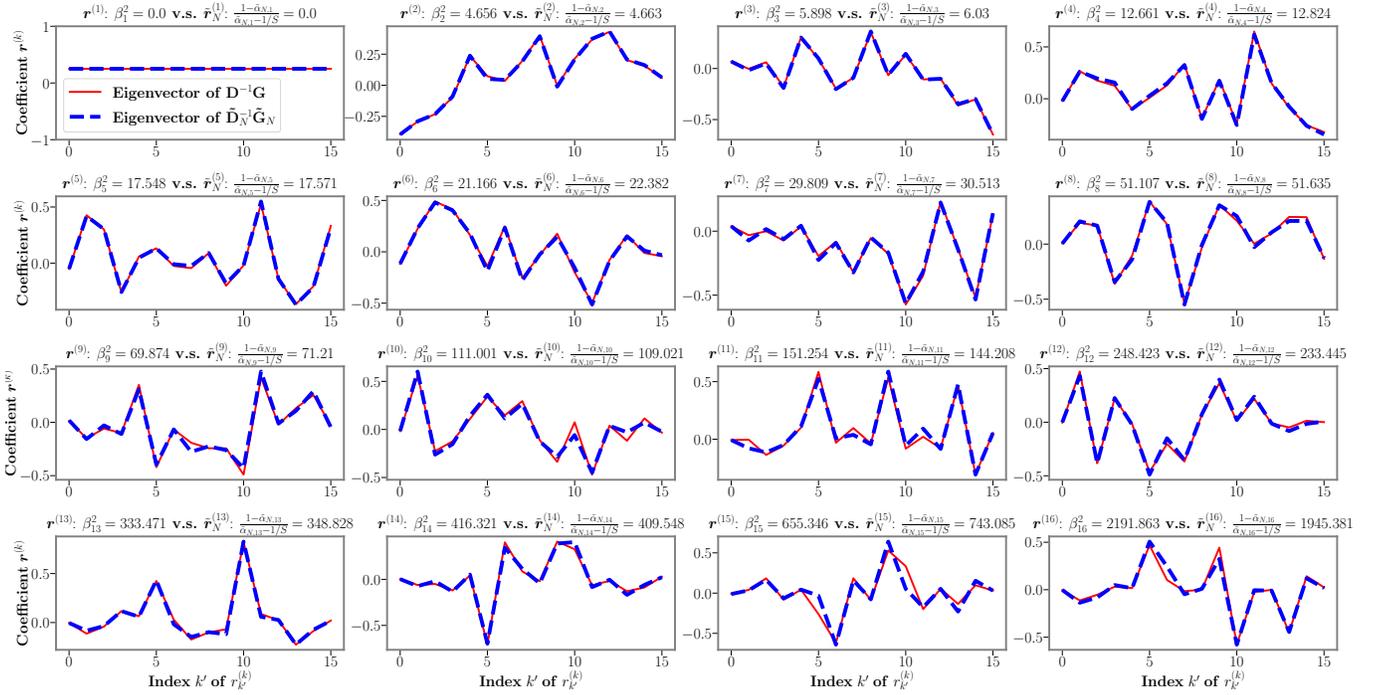}
    \caption{Estimating noise-to-signal ratio eigenvalues and corresponding eigentask coefficients under finite statistics ($N=300, \NS=1000$) in a 4-qubit H-encoding system, and comparison with theoretical value for $N\to\infty,\NS\to\infty$. }
    \label{fig:Tilde_Renormalization}
\end{figure}

\subsection{Gram matrix-free construction to approximate eigentasks and NSR eigenvalues}
\label{sec:GramMatrix-free}


If we consider Eq.\,(\ref{eq:DinvGeigenApprox}) and multiply through by $\mathbf{D}_N^{-\frac{1}{2}}$, the resulting equation can be written as an equivalent eigenproblem, 
\begin{align}
    \frac{1}{N} \widetilde{\mathbf{D}}_N^{-\frac{1}{2}} \regmat^T \regmat \widetilde{\mathbf{D}}_N^{-\frac{1}{2}} \left(\widetilde{\mathbf{D}}_N^{\frac{1}{2}} \tilde{\bm{r}}_N^{(k)}\right) = \tilde{\alpha}_{N,k}\left(\widetilde{\mathbf{D}}_N^{-\frac{1}{2}} \tilde{\bm{r}}_{N}^{(k)}\right) 
\end{align}
where we have also written $\widetilde{\gr}_N = \frac{1}{N}\widetilde{\mathbf{F}}_N^T\widetilde{\mathbf{F}}_N$ as in the previous section. Note that as written above, the eigenproblem is entirely equivalent to obtaining the singular value decomposition of the matrix $\frac{1}{\sqrt{N}} \widetilde{\mathbf{D}}_N^{-\frac{1}{2}} \regmat^T$. This particular normalization factor $\frac{1}{\sqrt{N}} \widetilde{\mathbf{D}}_N^{-\frac{1}{2}}$ is different from the standard z-score of principal components analysis. To obtain the combination coefficients $\bm{r}^{(k)}$, let $\bm{t}^{(k)} \in \mathbb{R}^K$ be the left singular vector of $\frac{1}{\sqrt{N}} \widetilde{\mathbf{D}}_N^{-\frac{1}{2}} \regmat^T$ (which is also the eigenvector of $\frac{1}{N} \widetilde{\mathbf{D}}_N^{-\frac{1}{2}} \regmat^T \regmat \widetilde{\mathbf{D}}_N^{-\frac{1}{2}} \approx \mathbf{D}^{-\frac{1}{2}} \widetilde{\gr} \mathbf{D}^{-\frac{1}{2}}$ in the large $N$ limit). Then $\bm{r}^{(k)} = \widetilde{\mathbf{D}}_N^{-\frac{1}{2}} \bm{t}^{(k)} \in \mathbb{R}^K$ can be treated as the combination prefactor of $\hat{M}_k$, to obtain the observables which correspond to the eigentasks. The merit of SVD analysis of $\frac{1}{\sqrt{N}} \widetilde{\mathbf{D}}_N^{-\frac{1}{2}} \regmat^T$ is that we only need to work with a $K$-by-$N$ matrix of features $\regmat$, which is numerically cheaper than further constructing a Gram matrix $\frac{1}{N} \regmat^T \regmat$. We will explore more about the usage of our technique in sense of PCA in Appendix \ref{app:PCA}.

\section{H-ansatz quantum systems: NSR spectra, expressive capacity, and eigentasks}
\label{app:H-ansztz}

In this section, we evaluate the EC for quantum systems described by the H-ansatz introduced in Appendix \ref{DetailsEncodings}, as an example of how EC can be efficiently computed for a variety of general quantum systems, and is not just restricted to parameterized quantum circuits.  The results of the analysis are compiled in Fig.\,\ref{fig:app_Features_and_Capacity}, and discussed below.


\begin{figure}
    \centering
    \includegraphics[width = 0.9\columnwidth]{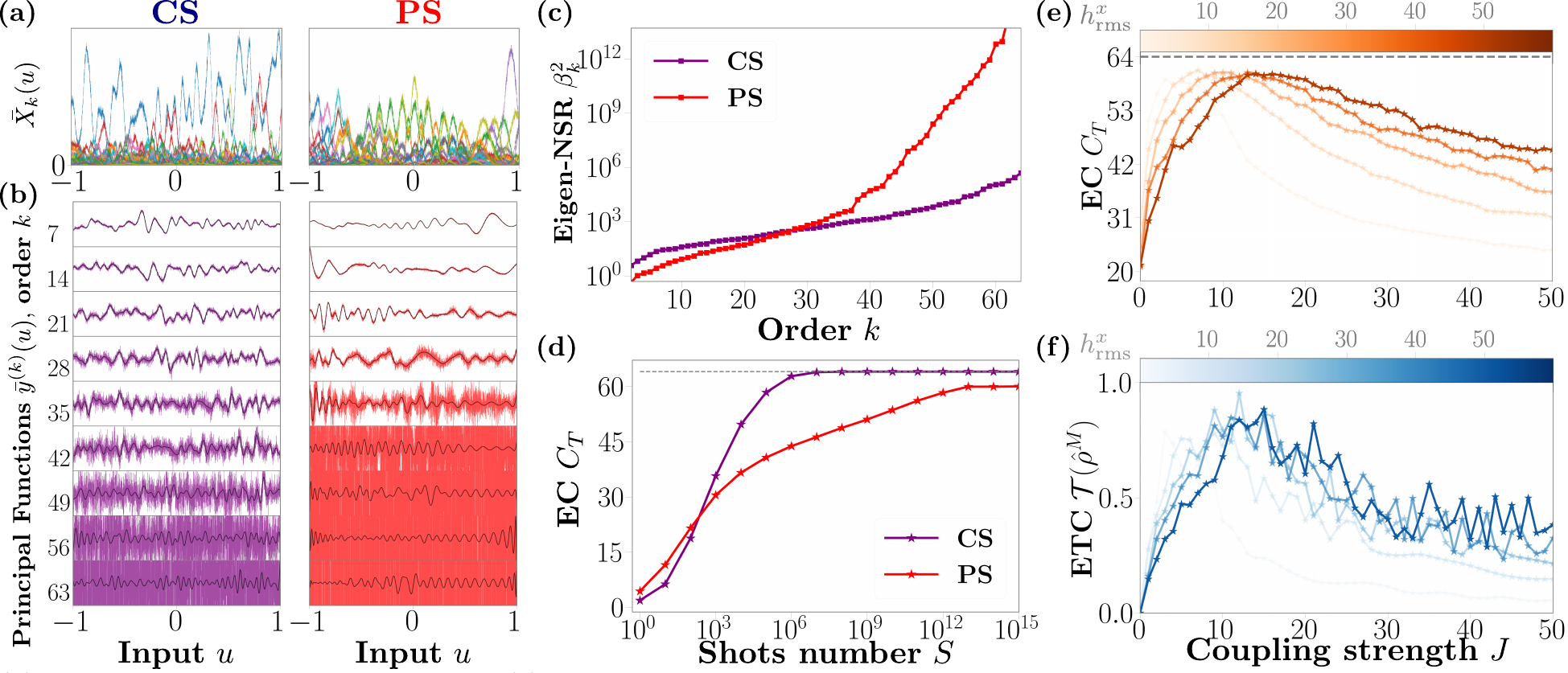}
    \caption{Eigen analysis in a $6$-qubit H-ansatz system (with $N = 5000$ and $\NS = 1000$) forming a 1D ring. The Hamiltonian parameters are chosen randomly with zero-mean and variance $(h^x_{\mathrm{rms}},  h^z_{\mathrm{rms}}, h^I_{\mathrm{rms}}) = (20, 5, 5)$, and $t = 5$ (See Appendix \ref{DetailsEncodings} for details). Coupling strength is uniformly $J \neq 0$ (correlated system) or $J=0$ (product system).
    (a) All $2^L=64$ noisy features $\bar{X}_k(u)$ and (b) noisy eigentasks $\bar{y}^{(k)}(u) = \bm{r}^{(k)}\cdot\bar{\bm{X}}(u)$ for selected $k$ from the features in (a), as well as their expected values $y^{(k)}(u) = \lim_{\NS\to\infty}\bar{y}^{(k)}(u)= \bm{r}^{(k)} \cdot \bm{x}(u)$ (black). 
    (c) Noise-to-signal ratio spectrum $\beta_k^2$ and (d) $\EC$ vs shots $S$ for both correlated system and product system encodings. 
    (e) $\EC$ at $\NS=10^5$ and (f) ETC, $\bar{\mathcal{T}}(\hat{\rho}^{M})$ in representative random $6$-qubit H-ansatz, as a function of coupling strength $J$. The peaks of capacity and correlation coincide, around $J \sim h_{\mathrm{rms}}^x$.
    }
    \label{fig:app_Features_and_Capacity}
\end{figure}


Fig.\,\ref{fig:app_Features_and_Capacity}(a) presents the set of features $\{\bar{X}_k(u)\}$ for typical $L=6$ qubit CS and PS at $\NS=1000$ with randomly chosen parameters (referred to as encodings, see caption). The resultant noisy eigentasks $\{\bar{y}^{(k)}(u)\}$ and NSR spectra $\{\beta_k^2\}$ extracted via the eigenvalue analysis are shown in  Figs.\,\ref{fig:app_Features_and_Capacity}(b) and \ref{fig:app_Features_and_Capacity}(c) respectively. In the side-by-side comparison in Fig.\,\ref{fig:app_Features_and_Capacity}(b), we clearly see the $J=0$ ansatz transitioning to a regime with more noise at much lower $k$ than the $J \neq 0$ ansatz. This is reflected in Fig.\,\ref{fig:app_Features_and_Capacity}(c), the $\beta_k^2$ spectrum, having a much flatter slope for larger $k$ (note the plot is semilog). Finally, Fig.\,\ref{fig:app_Features_and_Capacity}(d) shows the EC of both systems as a function of $\NS$. EC rapidly rises for small $S$ for both systems, but the rise of the $J=0$ system is steeper. After a certain threshold in $S$, however, the CS grows more rapidly, approaching the upper bound $2^6=64$ with $\NS \sim 10^8$; in contrast, the PS has a significantly lower $\EC$.

For $J \to \infty$ we also have $\bar{\mathcal{T}}=0$ because $\hat{\rho}_0 = \ket{0}\!\bra{0}^{\otimes L}$ is an eigenstate of the encoding ($\hat{\rho}(u)=\hat{\rho}_0$). This implies there must be a peak at some intermediate $J$, which for both EC and ETC occurs when the coupling is proportional to the transverse field $J \sim h^x$. 

 Our results elucidate the same kind of improvement, as can be observed when we consider how the EC $C$ changes with $J$, and compare it to the total correlation ETC $\bar{\mathcal{T}}$,  as shown in Fig.\,\ref{fig:app_Features_and_Capacity}(f). For $J \to 0$ we have a PS with $\bar{\mathcal{T}}=0$, whereas in the $J \to \infty$ we also have $\bar{\mathcal{T}}=0$ because $\hat{\rho}_0 = \ket{0}\!\bra{0}^{\otimes L}$ is an eigenstate of the encoding ($\hat{\rho}(u)=\hat{\rho}_0$).  This implies there must be a peak at some intermediate $J$, which for both EC and ETC occurs when the coupling is proportional to the transverse field $J \sim h^x$.  At finite $S$, increased ETC is directly related to a higher EC.

Another interesting aspect is the clear trend seen in the maximization of EC around $J \sim h^x_{\mathrm{rms}}$ for various $h^x_{\mathrm{rms}}$, possibly hinting at the role of increased correlation around the MBL phase transition in random spin systems~\cite{martinez2021dynamical}. This trend is consistent with results in quantum metrology -- in general, the SNR obtained from averaging $L$ uncorrelated probes scales as $1/\sqrt{L}$. This scaling can become favorable in the presence of quantum correlation and other non-classical correlations, in which case the scaling of the SNR can show up to a quadratic improvement $1/L$~\cite{Giovannetti2006}. For even larger $J$, we find that $\hat{\rho}(u)\to\hat{\rho}_0 = \ket{0}\!\bra{0}^{\otimes L}$, which clearly reduces $\ETC$, but also $\EC$ as the quantum system state becomes $u$-independent.


\section{Scaling with quantum system size}
\label{app:syssize}

An important question in quantum machine learning applications is the possible advantage of using larger quantum systems for information processing. In this section, we present preliminary results of scaling with quantum system size. The left panel of Fig.\,\ref{fig:Qubit_Scaling} shows EC vs $L$ at select $S$ values for H-ansatz, while the right panel shows two encodings in the C-ansatz device, as well as their noisy simulations. In both plots, the dashed line indicates the $S\to\infty$ result $\EC=2^L$. We see that the EC increases when adding more qubits into the Ising chain for the H-ansatz, or when increasing the number of circuit qubits $L$ for the C-ansatz. Note, however, that at any finite $S$ the noise-constrained EC falls off the exponential bound for $\NS\to\infty$. The dropoff is particularly severe for the IBMQ device, where we are limited to just $S\sim 10^4$, which significantly suppresses the EC even for $L=7$ qubits. Note, however, that even if one is well below $\EC=2^L$ due to this finite sampling constraint, increasing the dimension of the quantum system is always an effective way to increase the EC, particularly when compared to the logarithmic growth with $S$ of Fig.\,\ref{fig:Genc1} of Main Text. 

\begin{figure}[h]
    \centering
    \includegraphics[width=0.8\textwidth]{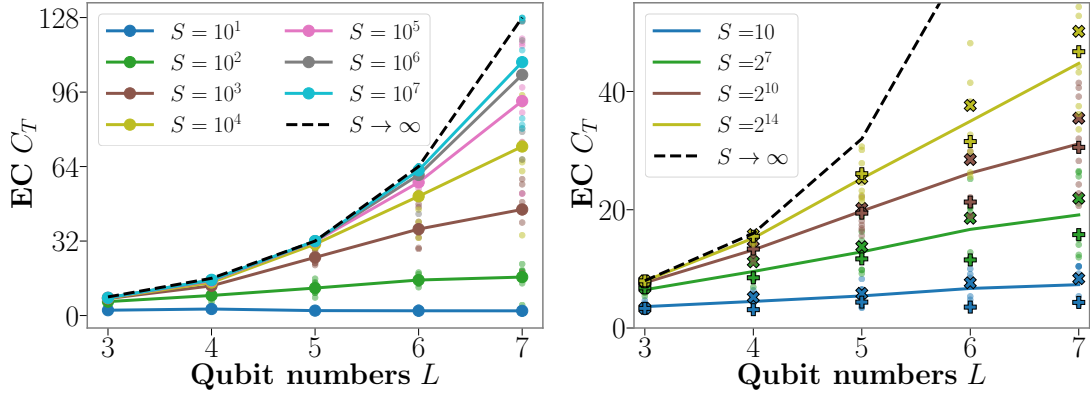}
    \caption{(a) H-ansatz and (b) C-ansatz at finite $S$ as a function of qubit number $L$. Various colours indicate different $S$ values, with the $S\to \infty$ bound in dashed black. Individual noisy simulations are indicated in small and transparent dots, with their average as a thick line, and the expressive capacity of the C-ansatz device for encoding 1 and 2 are indicated with `$\times$' and `$+$' respectively. }
    \label{fig:Qubit_Scaling}
\end{figure}


\section{Analytic solution to the quantum 2-design expressive capacity}
\label{app:2designsol}


\rev{The scaling of expressive capacity with system size in general is hard to quantify, and is likely best approached with a numerical or experimental study for a specific quantum model, as done in Fig.\,\ref{fig:Qubit_Scaling}.  However, we can analytically solve for the EC of a class of quantum models: \textit{$2$-design parametric quantum circuits} $\{ p(\bm{u}) \dd \bm{u}, \hat{U} (\bm{\theta} ; \bm{u}) \}$.  We clarify that we are referring here to systems with specific parameters $\bm{\theta}$ which result in 2-designs with respect to the input distribution $p(\bm{u})$; the ensemble average is taken with respect to  inputs $u$.  Quantum literature \cite{Holmes2022}  often refers to general ans\"atze which form $2$-designs with respect to parameters $\bm{\theta}$ instead, which is not what we are considering here.}

\rev{To be more specific, an ensemble $\{ p(\bm{u}) \dd \bm{u}, \hat{U} (\bm{\theta} ; \bm{u}) \}$ is a $2$-design if the following two quantum channels, defined on \textit{any} $2L$-qubit state, $\hat{\rho}$ are equal
\begin{equation}
    \mathcal{C} (\hat{\rho}) = \int \hat{U} (\bm{\theta} ; \bm{u})^{\otimes 2} \hat{\rho}_0 (\hat{U} (\bm{\theta} ; \bm{u})^{\dagger})^{\otimes 2} p (\bm{u}) \dd \bm{u} = \int \hat{U}^{\otimes 2} \hat{\rho}_0 (\hat{U}^{\dagger})^{\otimes 2} \dd \mu_H (\hat{U}) .
\end{equation}
where $\mu_{H}$ is the uniform (Haar) measure. We can verify that all information in the Gram matrix is explicitly contained in the elements of $\mathcal{C} (\hat{\rho}_0 \otimes \hat{\rho}_0)$. To be more specific,
\begin{align}
    & \bra{\bm{b}_{k_1}, \bm{b}_{k_2}} \mathcal{C} (\hat{\rho}_0 \otimes \hat{\rho}_0) \ket{\bm{b}_{k_1}, \bm{b}_{k_2}} \nonumber\\
    =~& \bra{\bm{b}_{k_1}, \bm{b}_{k_2}} \left( \int (\hat{U} (\bm{\theta} ; \bm{u}) \otimes \hat{U} (\bm{\theta} ; \bm{u})) \ket{\bm{b}_0, \bm{b}_0} \bra{\bm{b}_0, \bm{b}_0} (\hat{U} (\bm{\theta} ; \bm{u})^{\dagger} \otimes \hat{U} (\bm{\theta} ; \bm{u})^{\dagger}) p (\bm{u}) \dd \bm{u} \right) \ket{\bm{b}_{k_1}, \bm{b}_{k_2}} \nonumber\\
    =~& \int \left| \bra{\bm{b}_{k_1}} \hat{U} (\bm{\theta} ; \bm{u}) \ket{\bm{b}_0} \right|^2 \cdot \left| \bra{\bm{b}_{k_2}} \hat{U} (\bm{\theta} ; \bm{u}) \ket{\bm{b}_0} \right|^2 p (\bm{u}) \dd \bm{u} \nonumber\\
    =~& \int x_{k_1} (\bm{u}) x_{k_2} (\bm{u}) p (\bm{u}) \dd \bm{u} = (\mathbf{G})_{k_1 k_2} . 
\end{align}
However, $\mathcal{C} (\rho_0) = \int U^{\otimes 2} (\hat{\rho}_0 \otimes \hat{\rho}_0) (U^{\dagger})^{\otimes 2} \dd \mu_H (U)$ implies that we can compute the Gram matrix by instead integrating over the Haar measure \cite{Puchaa2017}:
\begin{align}
  (\mathbf{G})_{k_1 k_2} = \int | U_{0, k_1} |^2 | U_{0, k_2} |^2 \dd \mu_H (U) = \left\{\begin{array}{ll}
    \frac{2}{K (K + 1)}, & \text{if } k_1 = k_2,\\
    \frac{1}{K (K + 1)}, & \text{if } k_1 \neq k_2 .
  \end{array}\right. 
\end{align}
Then the corresponding second-order matrix $\mathbf{D}$ is given by
\begin{equation}
    (\mathbf{D})_{k k} = \frac{2}{K (K + 1)} + (K - 1) \times \frac{1}{K (K + 1)} = \frac{1}{K} .
\end{equation}
It is self-consistent that the matrix $\mathbf{D}= \mathrm{diag} \left( \frac{1}{K}, \frac{1}{K}, \cdots, \frac{1}{K} \right)$ obeys the normalization condition $\mathrm{Tr} (\mathbf{D}) = K \cdot \frac{1}{K} = 1$. Then we can solve the eigenvalues $\{ \alpha_k \}_{k \in [K]}$ of random walk matrix $(\mathbf{D}^{- 1} \mathbf{G})_{k_1 k_2} = \frac{1}{K + 1} (1+\delta_{k_1 k_2})$. It gives
\begin{equation}
     \alpha_k = \frac{1}{K + 1}(1 + K \delta_{k0}).
\end{equation}
Furthermore, we use $\alpha_k = \frac{1}{1 + \beta_k^2}$ or $\beta^2_k = \frac{1}{\alpha_k} - 1$ to compute the eigen-NSR:
\begin{equation}
    (\beta^2_0, \beta^2_1, \beta^2_2, \cdots, \beta^2_{K - 2}, \beta^2_{K - 1}) = (0, K, K, \cdots, K, K) .
\end{equation}
Then the expressive capacity of any 2-design system is given by
\begin{equation}
    C_T = 1 + \frac{K - 1}{1 + \frac{K}{S}} = K \times \frac{1 + \frac{1}{S}}{1 + \frac{K}{S}} = 2^L \times \frac{S + 1}{S + 2^L} .
\end{equation}
}


\section{Quantum correlation metrics}
\label{app:QCM}

There is no one standard metric to quantify correlation in a many-body state. The metric we introduce here, the \textit{quantum total correlation}, is a quantity inspired by the classical total correlation of $L$ random variables $(b_1, \cdots, b_L)$, that is $\sum_{l = 1}^L \mathrm{H} (b_l) - \mathrm{H} (b_1 , \cdots, b_L )$. Using chain rule of Shannon entropy $\mathrm{H} (b_1, b_2, \cdots, b_L) = \mathrm{H} (b_1) + \mathrm{H} (b_2 | b_1 ) + \cdots + \mathrm{H} (b_L | b_1, b_2, \cdots, b_{L - 1} )$
\begin{align}
    & \sum_{l = 2}^L \mathrm{H} (b_l) - \mathrm{H} (b_1, b_2, \cdots, b_L) = \sum_{l = 1}^L \mathrm{H} (b_l) - \sum_{l = 1}^L \mathrm{H} (b_l | b_1, b_2, \cdots, b_{l - 1} ) = \sum_{l = 2}^L \mathrm{I}(b_1, \cdots, b_{l - 1} ; b_l) \in [0, L-1], 
\end{align}
we can see that the classical total correlation tells us how a set of random variables reveals information of each other. Similarly, quantum total correlation can be defined as \cite{Vedral2002, Modi2010}
\begin{align}
    \mathcal{T} ( \hat{\rho} ) = \sum_{l = 1}^L \mathrm{S} ( \hat{\rho}_l ) - \mathrm{S} ( \hat{\rho} )
\end{align}
where $\mathrm{S}$ is von Neumann entropy and $\hat{\rho}_l := \mathrm{Tr}_{[L] \backslash \{ l \}} \left\{ \hat{\rho} \right\} $ is the subsystem state at qubit $l$. Due to the subadditivity of von-Neumann entropy $\sum_{l = 1}^L \mathrm{S} ( \hat{\rho}_l )  \geq \mathrm{S} ( \hat{\rho} )$, we conclude that the quantum total correlation is non-negative, and is zero iff the state $\hat{\rho} = \bigotimes_{l=1}^{L} \hat{\rho}_l$ is a product state. 

In this paper's measurement scheme, the specific readout POVMs are the projectors onto the computational states $\{ \ket{\boldsymbol{b}_k} \bra{\boldsymbol{b}_k} \}_{k \in [K]}$. Thus, we are in particular interested in analyzing the post-measurement state $\hat{\rho}^M (u) = \sum_k \rho_{k k} (u)  \ket{\boldsymbol{b}_k} \bra{\boldsymbol{b}_k}$ whose subsystems are correspondingly in states $\hat{\rho}_l^M (u) = \mathrm{Tr}_{[L] \backslash \{ l \}} \left\{ \hat{\rho}^M (u) \right\}$. We compute the average or \textit{expected} quantum total correlation over the input domain $u$ with respect to the input probability distribution $p(u)$: 
\begin{align}
    \bar{\mathcal{T}}\! \left( \hat{\rho}^M \right) = \mathbb{E}_u \left[ \sum_{l = 1}^L \mathrm{S} ( \hat{\rho}_l^M (u) ) - \mathrm{S} (
    \hat{\rho}^M (u) ) \right] =\mathbb{E}_u \left[ \sum_{l = 1}^L \mathrm{H} (b_l (u)) - \mathrm{H} (b_1 (u), \cdots, b_L (u)) \right]
\end{align}
where the second equality comes from the diagonal nature of post-measurement state which reduces the quantum total correlation to a normal classical total correlation.

The post-measurement quantum total correlation always reaches its maximum $L-1$ when the diagonal terms of the state is a GHZ-type state. Also as a comparison, for a $W$-state $\ket{W} = \frac{1}{\sqrt{L}} \left( \ket{10 \cdots 0} + \ket{01 \cdots 0} + \cdots + \ket{00 \cdots 1} \right)$, then post-measurement quantum total correlation $\mathrm{T} ( \ket{W} )$ is
\begin{align}
     L \left( - \left( \frac{1}{L} \right) \log_2 \left( \frac{1}{L} \right) - \left( \frac{L - 1}{L} \right) \log_2 \left( \frac{L - 1}{L} \right) \right) - L \left( - \left( \frac{1}{L} \right) \log_2 \left( \frac{1}{L} \right) \right) = (L - 1) \log_2 \left( \frac{L}{L - 1} \right) .
\end{align}
which is upper bounded by $\lim_{L \rightarrow \infty} \mathcal{T} ( \ket{W} ) = \frac{1}{\ln (2)} \approx 1.443$.


\section{Guidance from EC theory: principal component analysis with respect to quantum noise}
\label{app:PCA}

Our proposed capacity spectrum analysis has another significant benefit: it provides a natural truncation scale for eigentasks. In machine learning theory, the technique of projection of a high-dimensional data to a subspae of reduced dimensionality is called \textit{principal component analysis}. Within the computing architecture we are discussing, we are interested in carrying out a PCA in the function space. 
 More specifically, consider a given function $f(u)$, we hope to find $K'$ functions $\{G^{(k)}(u)\}_{k \in [K']}$ where $G^{(k)}(u)=\sum_{k' = 0}^{K-1} g^{(k)}_{k'} x_{k'}(u)$ lies in the space spanned by measured features $G^{(k)}(u) \in \mathrm{Span}\{\bm{x}\}$, such that the relative mean square error 
\begin{align}
    \min_{\bm{W}} \frac{\mathbb{E}_u \! \left[\left|f - \sum_{k=1}^{K'} W_k \left(\sum_{k' = 0}^{K-1} g^{(k)}_{k'} \bar{X}_{k'}\right) \right|^2\right]}{\mathbb{E}_u[|f|^2]}
\end{align}
is much smaller as possible. According to Appendix \ref{sec:Information_capacity_saturation}, the solution to $\{\bm{g}^{(k)}\}_{k \in [K']}$ is exactly $\bm{g}^{(k)} = \bm{r}^{(k)}$. Fig.\,\ref{fig:PCA} supplies a concrete example of fitting linear function $f(u)=u$, by setting $K'=40$ in a $6$-qubit system (and thus $K=64$). 
\begin{figure}[h]
    \centering
    \includegraphics[width=0.8\columnwidth]{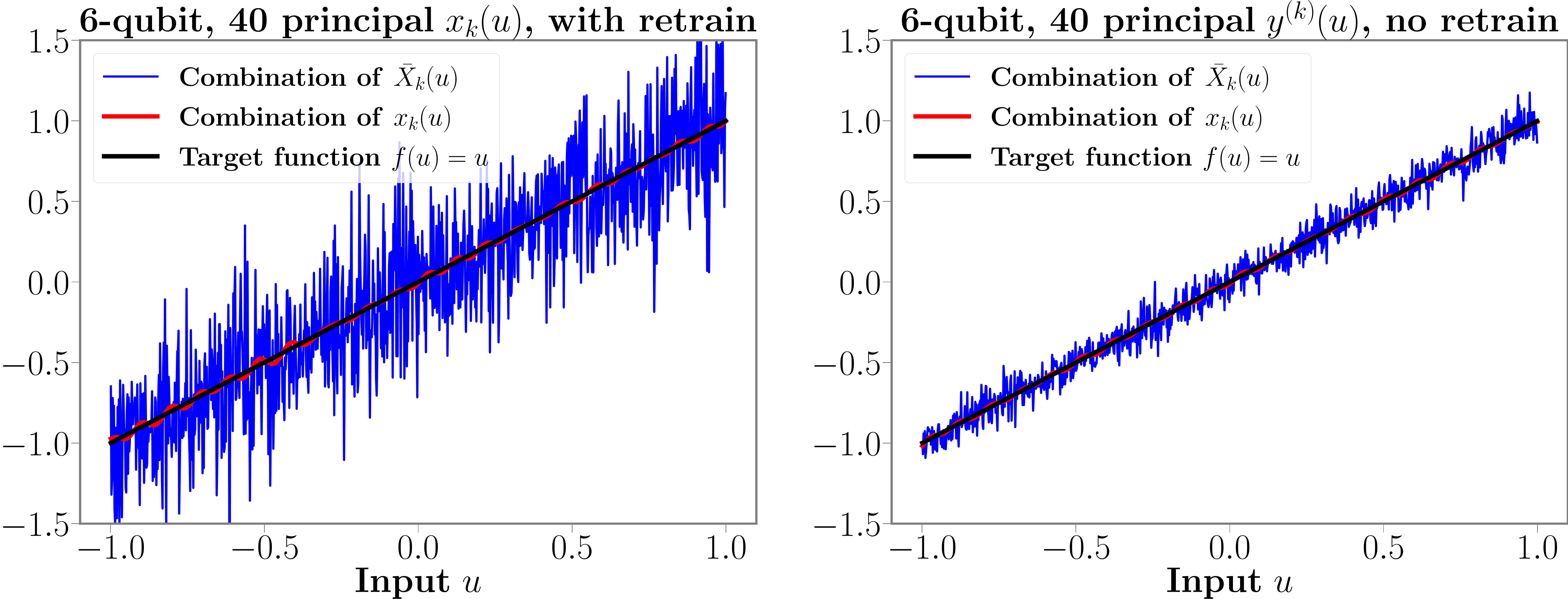}
    \caption{Projection onto $40$-dimensional space spanned by $40$ principal $x_k(u)$ vs. spanned by first $40$ eigentasks $y^{(k)}$, in a $6$-qubit H-encoding system. The number of shots is fixed as $\NS=5000$. }
    \label{fig:PCA}
\end{figure}

Fig.\,\ref{fig:PCA}(a) shows the projection onto the space spanned by the dominant $40$ readout features. Here, by ``dominant" we mean one can first train by least square regression to get an output weight $\bm{w} \in \mathbb{R}^{K}$, and then select corresponding $w_k$ with the leading $K'$ largest $w^2_k \cdot \mathbb{E}_u[|x_k|^2]$. Then we need to use these $K'$ features to retrain and obtain a new output weight $\bm{w}' \in \mathbb{R}^{K'}$. In such particular example, $\bm{g}^{(k)}$ are some one-hot vectors where the index of $1$ are chosen by the sorting $K'$ largest $w^2_k \cdot \mathbb{E}_u[|x_k|^2]$ as we described before. We can compare the relative mean square error with the case of $\bm{g}^{(k)} = \bm{r}^{(k)}$, the eigentasks. As illustrated in Fig.\,\ref{fig:PCA}(b) the latter is able to achieve an approximation of the desired function (here a linear function) with a decidedly smaller relative mean square error. 

One important question is: what would be an appropriate choice of $K'$ in practice? In Appendix \ref{app:Spectral_finite_statistics} we claim that it is determined by the set of eigentasks for which $\beta_k^2/S < 1$, those for which the signal is larger than the noise. Namely we should define the cut-off $K_{c}(S)$ such that 
\begin{align}
    K_{c}(S) = \max_{\beta_{k}^2 < \NS} k. 
\end{align}
Based on this observation, we can further explore the trend of $K_{c}(S)$ when qubit number $L$ is scaled. As we show in the main text, the eigen-NSR spectra grow much slower when $L$ increases. Then the quantum system is able to provide much more eigentasks with more signal than noise. Fig.\,\ref{fig:PCA_L}(a) shows spectrum in H-encoding quantum system with sizes ranging from $L=3$ to  $L=8$ with fixed hyperparameters. Notice that number of shots $\NS=5000$ here is not a large number, which means that we cannot sample enough shots so that features converge to their mean values in the $2^8=256$ dimensional Hilbert space. But applying eigentasks analysis in this example still shows a fast decay of relative error $\min_{\bm{W}} \mathbb{E}_u  [|f - \sum_{k=0}^{K_c(S)} W_k \bar{y}^{(k)} |^2]/\mathbb{E}_u[|f|^2]$ until the fitting accuracy saturates at $L=8$.  

\begin{figure}[h]
    \centering
    \includegraphics[width=1.0\columnwidth]{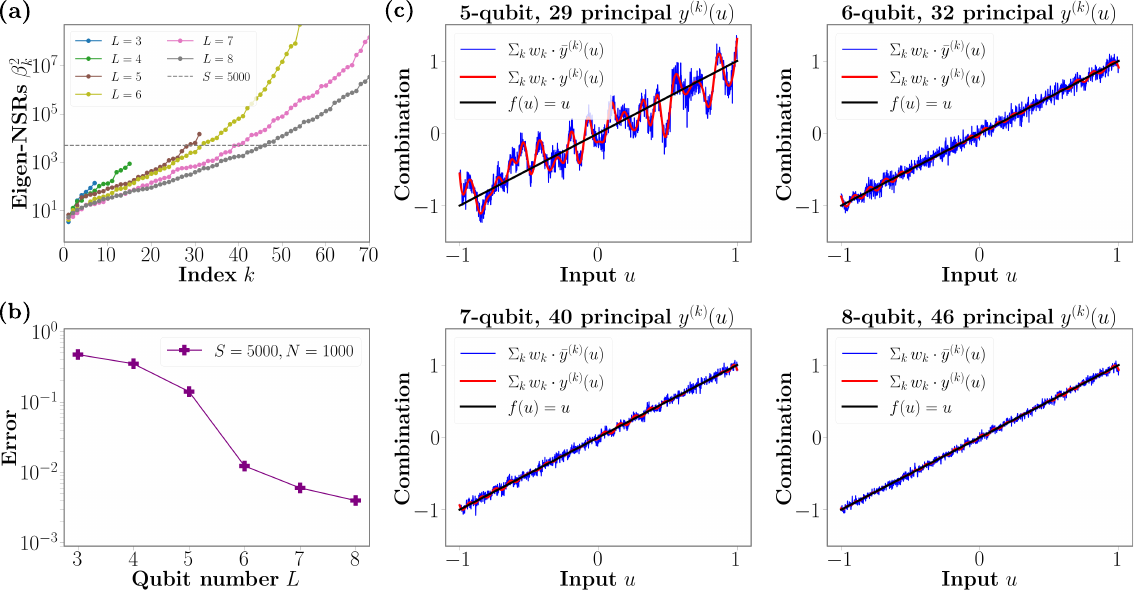}
    \caption{PCA for different CS H-encoding system size $L=3,4,5,6,7,8$ with fixed hyperparameters and $S=5000$. (a) Eigen-noise-to-signal ratios spectrum of different sized system. (b) Relative error $\min_{\bm{W}} \mathbb{E}_u  [|f - \sum_{k=1}^{K_c} W_k \bar{y}^{(k)} |^2]/\mathbb{E}_u[|f|^2]$ for fitting $f(u)=u$, where $K_c$ can be read out from (a). (c) Combination of $K_c$ eigentasks $\sum_{k=0}^{K_c(S)} w_k y^{(k)}(u)$ and noisy eigentasks $\sum_{k=0}^{K_c(S)} w_k \bar{y}^{(k)}(u)$ in $L=5,6,7,8$ qubits system.}
    \label{fig:PCA_L}
\end{figure}


\section{Quantum-noise-PCA in classification problem}
\label{sec:Single_step_QRC}

The highly nonlinear readout feature $x_k(u)$ should have Taylor expansion $x_k(u) = \sum_j^{\infty} (\mathbf{T})_{kj} u^j$. Such complicated functions will span a certain functional space. One fundamental question is what the limit of approximation ability is based on the architecture we proposed. 
Hereby we first show that this architecture \textit{under infinite sampling} is capable of approximating \textit{any} continuous function on the domain $[-1, 1]$ to arbitrary precision. 
Furthermore, the linearity of quantum moment readout and complexity of quantum evolution will help us to understand why such a quantum system has capability to approximate a highly nonlinear function, under finite and bounded computational resources. Exploring the capacity for function approximation under finite measurement resources, as is done in the main text and Appendix \ref{sec:Information_capacity_saturation}, highlights the fundamental limitations placed by quantum noise on computation using the reservoir computing approach. 

\subsection{Universal Function Approximation}
\label{FuncApxUniv}

A very general question is that what type of functions  such a single-step quantum evolution can approximate. One conclusion which can be drawn is the \textit{universal function approximation} property. That is, give any continuous function from space of continuous functions on domain $[-1, 1]$, \textit{i.e.} $\phi \in \mathscr{C}([-1, 1], \mathbb{R})$, for any given error $\varepsilon > 0$, there always exists a function $\varphi(u) = \boldsymbol{w} \cdot \boldsymbol{x}(u)$ such that 
\begin{align}
    |\varphi(u) - \phi(u)| \leq \varepsilon
\end{align}
for any input $u \in [-1,1]$. The proof employs the well-known Stone-Weierstrass theorem. For our particular architecture, $D=[-1, 1]$ is obviously a compact space, while point-separation can also be trivially fulfilled by a single qubit system ($L=1$). The subalgebra structure of the function family generated by quantum systems is automatically satisfied in family of all product systems, as long as we use a \textit{Walsh-Hadamard transformation} to convert from the quantum probability $\bm{x}(u)$ to the many-body Pauli-$z$ products $\{\langle \prod_i \hat{\sigma}^z_{l}: l \in B \rangle_{\hat{\rho}}\}$ for all $B \subseteq [L]$. 
\begin{figure}[tbp]
    \centering
    \includegraphics[width = 0.80\columnwidth]{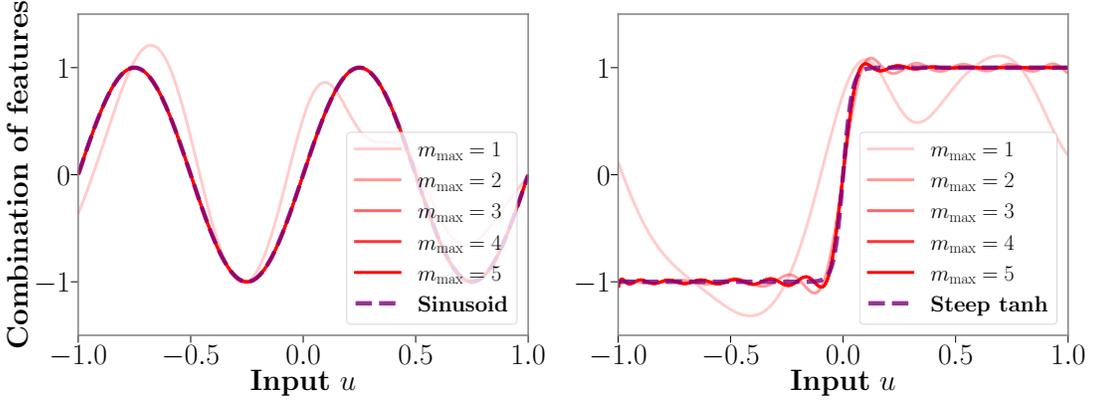}
    \caption{Function approximation by using $y = \sum_{k=0}^{K-1} w_k x_k(u)$ (solid red lines) to approximate sine function and steep tanh function (dashed purple lines) in a 5-qubit quantum annealing system, where $K_{\rm eff} = \sum_{m=0}^{m_{\rm max}}\big(^L_{m}\big)$ depends on different quantum moment thresholds $m_{\rm max} = 1,2,3,4,5$. The hyperparameters are $(J_{\rm max}; \bar{h}^x, h^x_{\mathrm{rms}}; \bar{h}^{I}, h^{I}_{\mathrm{rms}}) = (1; 3, 1; 5, 2)$ in unit $1/t$ and no $h^z$ field. This simulation shows that for some simple functions, it is sufficient to merely use lower order moments, \textit{e.g.}, $m_{\rm max}=2$ in sine function and $m_{\rm max}=3$ in steep tanh function. }
    \label{fig:Universality_tanh}
\end{figure}

\subsection{1D classification as function approximation for noiseless measured features}
In this section, we will show how the universal function approximation property of the architecture described in Appendix \ref{FuncApxUniv} enables it to perform -- among others -- paradigmatic machine learning tasks such as classification. 
\begin{figure}[tbp]
    \centering
    \includegraphics[width = 0.8\columnwidth]{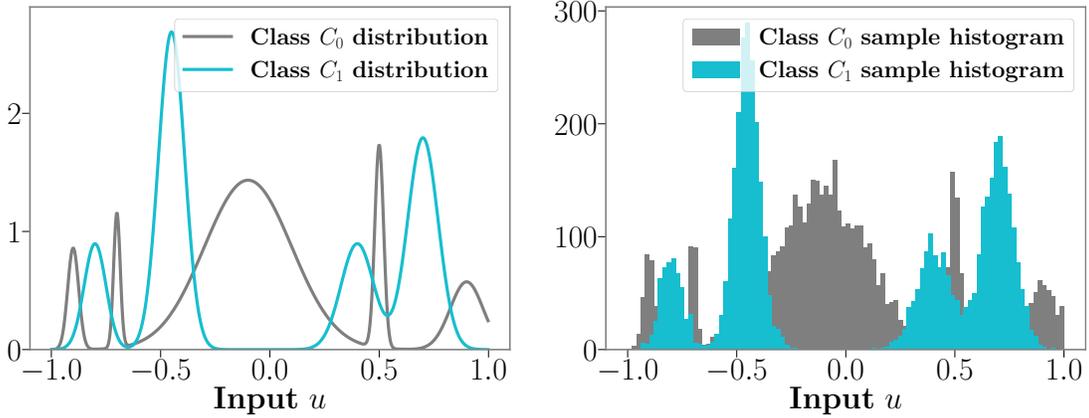}
    \caption{(Left) Distribution $p_0(u)$ and $p_1(u)$ for classes $C_0$ and $C_1$, respectively. (Right) The histogram of $C_0$ and $C_1$. Each class contains $5000$ samples. }
    \label{fig:distribution_histogram}
\end{figure}

Suppose two classes $C_0$ and $C_1$ of samples, each of which is generated from distributions $p_0(u)$ and $p_1(u)$ respectively. The probability of occurrence of $C_0$ and $C_1$ are both $50\%$, and we simply let each class equally contain $5000$ samples and thus $N = 10000$ samples in total. Both distribution are artificially defined by summing several Gaussian distributions with different amplitudes and widths together. Domain of both distributions are restricted in $[-1, 1]$ and both distributions are also normalized. Due to the overlap of two distributions, there is some theoretical maximal classical accuracy to distribution whether a given $u$ belongs to either $C_0$ or $C_1$. 

During the training, we feed each sample $u^{(n)}$ (belonging to class $C_{c^{(n)}}$) into a $5$-qubit quantum system. The quantum system will be read out with $K_{\rm eff} = \sum_{m=0}^{m_{\rm max}}\big(^L_{m}\big)$ different features $\{x_k(u^{(n)})\}_{k \in [K_{\rm eff}]}$. Then features of $N$ sample forms the regressor matrix. According to the standard supervised learning procedure, we simply train based on $(\boldsymbol{x}(u^{(n)}), c^{(n)})$ by logistics regression where one should minimize the cross-entropy loss 
\begin{align}
    \mathscr{L} (\boldsymbol{W}) =~ \frac{1}{N} \sum_{n = 1}^{N} \bigg[- c^{(n)} \mathrm{log}\!\left( \sigma (\boldsymbol{W}\cdot\boldsymbol{x}(u^{(n)})) \right) - \left( 1 - c^{(n)} \right) \mathrm{log}\!\left( 1 - \sigma (\boldsymbol{W}\cdot\boldsymbol{x}(u^{(n)})) \right)\bigg]
    \label{eq:cross-entropy}
\end{align}
where $\sigma$ is the sigmoid function $\sigma(y) = \frac{1}{1+e^{-y}}$. A small $L_2$ penalty $\lambda \|\boldsymbol{W} \|^2$ (where $\lambda=10^{-6}$) is added to Eq.\,(\ref{eq:cross-entropy}) for preventing overfitting. The optimal $\bm{W}$ is then simply the set of weights that minimizes this cost function,
\begin{align}
    \bm{w} = {\rm argmin}_{\bm{W}}~\{\mathscr{L} (\boldsymbol{W})\}
\end{align}

\begin{figure}[t]
    \centering
    \includegraphics[width = 0.8\columnwidth]{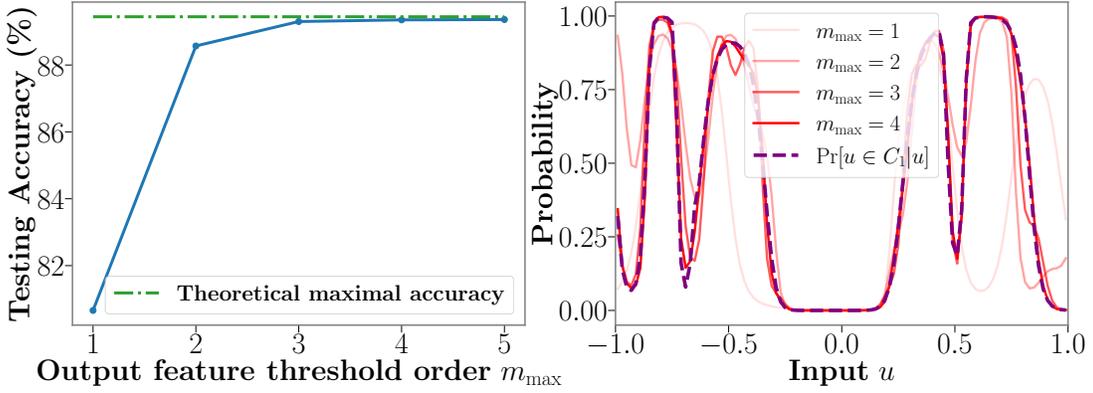}
    \caption{1D classification as function approximation in a $5$-qubit quantum system with full connectivity. The hyperparameters are $(J_{\mathrm{max}}; \bar{h}^x, h^x_{\mathrm{rms}}; \bar{h}^I, h^I_{\mathrm{rms}}) = (1; 3, 1; 8, 5)$ in unit $1/t$ and no $h^z$ field. (Left) Testing accuracy as a function highest order $m_{\rm max}$ of moment feature. (Right) Conditional distribution $\mathrm{Pr}[u \in C_1|u]$ (purple dashed line) vs. readout features $\sigma(\boldsymbol{w} \cdot \boldsymbol{x} (u))$ with $m_{\rm max} = 1,2,3,4$ (red solid line). $m_{\rm max}=4$ saturates the approximation accuracy. }
    \label{fig:Approximate_conditional_probability}
\end{figure}

We test the fidelity of learning the classification task by determining the accuracy of classification on a testing set formed by drawing $N = 10000$ new samples (independent of the training set) as a function of the order of output moments extracted, $m_{\rm max} = 1,2,3,4,5$, corresponding to reading out $K_{\rm eff}=6, 16, 26, 31, 32$ features respectively. The resulting testing accuracy is plotted in the left panel of Fig.\,\ref{fig:Approximate_conditional_probability}). We see that the testing accuracy converges to the theoretical maximal accuracy (dashed green) with increase in readout features.

Importantly, one can show that this improvement in learning performance coincides with training of optimal weights $\bm{w}$ such that the QRC is able to approximate the conditional distribution $\mathrm{Pr}[u \in C_1 | u]$ of the two classes with increasing accuracy (lower error). To verify this, we first numerically compute all $K=32$ readout feature functions $\boldsymbol{x}(u)$ of the system, by sweeping $500$ equidistant values of $u \in [-1, 1]$. Effectively learning the conditional distribution means that $\sigma(\boldsymbol{w} \cdot \boldsymbol{x}(u)) \approx \mathrm{Pr}[u \in C_1 | u]$. It is equivalent to use $\boldsymbol{w} \cdot \boldsymbol{x}(u)$ to approximate the following function: 
\begin{equation}
    \boldsymbol{w} \cdot \boldsymbol{x}(u) \approx \sigma^{-1}(\mathrm{Pr}[u \in C_1 | u]). 
    \label{eq:classifier_approx}
\end{equation}
We therefore see that the function approximation universality property of the architecture discussed in Appendix \ref{FuncApxUniv} enables its use as a generic classifier.

\subsection{Solving classification problem by quantum-noise-PCA}

\begin{figure}[t]
    \centering
    \includegraphics[width = 0.8\columnwidth]{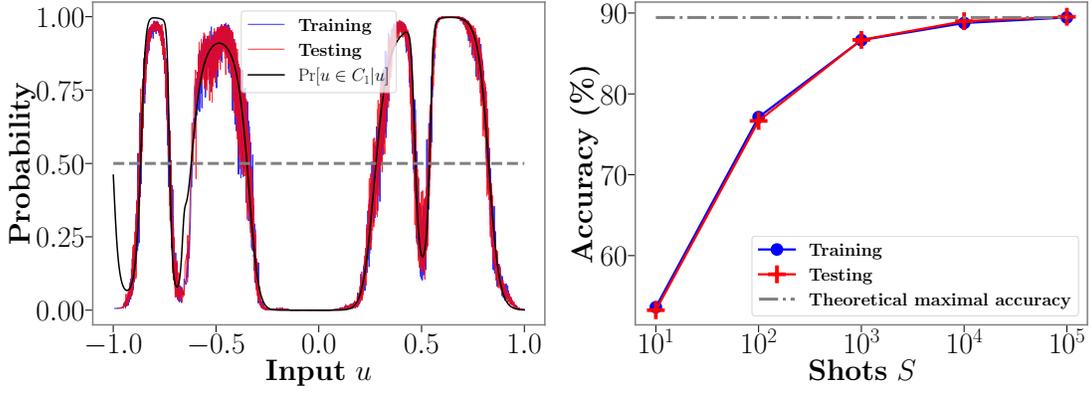}
    \caption{(Left) The linear combination with sigmoid activation, that is the stochastic function $\sigma\left(\sum_{k'=1}^{K_c(S)} w_{k', \mathrm{Train}} (\tilde{\bm{r}}_N^{(k)} \cdot \bar{\bm{X}}_{\mathrm{Train}})_{k'} \right)$ (blue line) and $\sigma\left(\sum_{k'=1}^{K_c(S)} w_{k', \mathrm{Train}} (\tilde{\bm{r}}_N^{(k)} \cdot \bar{\bm{X}}_{\mathrm{Test}})_{k'} \right)$ (red line), compared with the true conditional probability $\Pr[u \in C_1|u]$ (black line). (Right) Training accuracy and testing accuracy. They saturate the theoretical maximal accuracy as $S$ reaches $10^4 \sim 10^5$. Their agreement shows the quantum measurement noise serves well as a regularizer. }
    \label{fig:Classification_ES}
\end{figure}

Now we can solve the classification task above by using the quantum-noise princilpal component analysis we learn from capacity analysis. Suppose a physical system with $L=5$ qubits and ring connectivity, we choose the hyperparameter to be $J=2$, $ h^x_{\mathrm{rms}} = h^z_{\mathrm{rms}} = h^I_{\mathrm{rms}} = 5$ and $t=3$. In this H-encoding scheme, we can obtain $K=32$ measured features on each of $N=10^5$ samples $\{u^{(n)}\}$ ($5000$ in class $C_0$ and $5000$ in class $C_1$). We emphasize here that the underlying marginal distribution $p(u)$ is no longer uniform here, and it will make both $\{\beta_k^2\}$ and $\{\bm{r}^{(k)}\}$ very different. 

Given the number of shots $S \in [10^1, 10^5]$, we can still compute the empirical $\tilde{\bm{r}}_N^{(k)}$ and estimating $\beta_k^2$ by using the correction techniques we used in Appendix 
 \ref{app:Spectral_finite_statistics}. By comparing the estimated $(1 - \tilde{\alpha}_{N,k})/(\tilde{\alpha}_{N,k} - \frac{1}{\NS})$ and $S$, we can figure out the cutoff order $K_c(S)$ and combination coefficients $\tilde{\bm{r}}_N^{(k)}$, based on which we can define a set of observables
\begin{align}
    \hat{O}_k = \sum_{k' = 0}^{K-1} \tilde{\bm{r}}^{(k)}_{N,k'} \hat{M}_{k'} \quad k = 0, 1, \cdots, K_c(S).
\end{align}
It is equivalent to say, by measuring $\hat{O}_k$, we can effectively obtain eigentasks $\tilde{\bm{r}}_N^{(k)} \cdot \bar{\bm{X}}_{\mathrm{Train}}$. Then we can apply standard logistics regression on those eigentasks as we did in Eq.\,\ref{eq:cross-entropy}. The only difference is we no longer need any regularization term as penalty like $\lambda \|\boldsymbol{W} \|^2$. The training procedure eventual yield $\bm{w}_{\mathrm{Train}} \in \mathbb{R}^{K_c(S)}$, together with $\tilde{\bm{r}}_N^{(k)}$ and $K_c(S)$. 

Now we generate a totally new and independent set of $u$'s for testing purpose. By measuring $\hat{O}_k$, one get eigentasks $\tilde{\bm{r}}_N^{(k)} \cdot \bar{\bm{X}}_{\mathrm{Test}}$. By plugging $\bm{w}_{\mathrm{Train}} \in \mathbb{R}^{K_c(S)+1}$, together with $\tilde{\bm{r}}_N^{(k)}$ and $K_c(S)$ in training, we can achieve the testing accuracy. The agreement between training and testing accuracy show that the quantum measurement noise effectively works as a regularizer, and do a pretty good job (see Fig.\,\ref{fig:Classification_ES}).


\section{Finite sampling bound and uncertainty propagation}
\label{app:Uncertainty_propagation}
We conclude that the principle advantage brought about by correlation in this sections.  
There we observe that for certain inputs $u$ (that depend on the input encoding) the measurement of an CS when mapped into the moment space can generate distributions that can be highly anisotropic at finite $\NS$. While for PS these distributions are generally isotropic unless they are close to the boundaries of the output domain (when the encoding produces outputs that are eigenstates of the measurement basis). We observe that this trend is also present in the experimental system despite non-idealities. The origin of higher expressive capacity at large $\NS$ provided by ESs can be traced back to this basic feature. 
To be more specific, let $\hat{N}_k = \hat{\sigma}_{l_1}^z \hat{\sigma}_{l_2}^z \cdots \hat{\sigma}_{l_m}^z$, and $\bar{X}_{k} (u)$ be empirical mean based on $\NS$ sampling of Pauli-$z$ products. Notice that the variance of $\bar{X}_k$ now has an alternative fomr of
\begin{equation}
    \mathrm{Var} [\bar{X}_k] = \frac{1}{\NS} \left( \langle (\hat{\sigma}_{l_1}^z \hat{\sigma}_{l_2}^z \cdots \hat{\sigma}_{l_m}^z)^2 \rangle - \langle \hat{\sigma}_{l_1}^z \hat{\sigma}_{l_2}^z \cdots \hat{\sigma}_{l_m}^z \rangle^2 \right) = \frac{1}{\NS}(1 - x^2_k (u)) .
\end{equation}
Thus,
\begin{equation}
    \bar{X}_{k} (u) = x_k (u) + \delta_{k}(u) = x_k (u) + \frac{1}{\sqrt{\NS}} \zeta_{k}(u),
\end{equation}
where random sampling noise $\zeta_{k}(u) \approx \sqrt{1 - x_k^2 (u)} \epsilon$ and $\epsilon$ is a random fluctuation with variance $1$. For quantum moment readout, the amplitude of relative error is
\begin{equation}
    \left| \frac{\delta_{k}(u)}{x_k (u)} \right| \approx \sqrt{\frac{1 - x_k^2 (u)}{x_k^2 (u)}} \frac{1}{\sqrt{\NS}} \propto \frac{1}{\sqrt{\NS}} .
\end{equation}
For classical polynomial readout the amplitude of relative error is obtained by rule of uncertainty propagation
\begin{align}
  & \left| \frac{(x_{l_1} (u) + \delta_{l_1}) \cdots (x_{l_m} (u) + \delta_{l_m}) - x_{l_1} (u) \cdots x_{l_m} (u)}{x_{l_1} (u) \cdots x_{l_m} (u)} \right|
  \approx \left| \frac{\delta_{l_1}}{x_{l_1} (u)} + \cdots + \frac{\delta_{l_m}}{x_{l_m} (u)} \right| \nonumber\\
  \approx & \left( \sqrt{\frac{1 - x^2_{l_1} (u)}{x^2_{l_1} (u)}} + \cdots + \sqrt{\frac{1 - x^2_{l_m} (u)}{x^2_{l_m} (u)}} \right) \times \frac{1}{\sqrt{\NS}}
  \propto \, m \times \frac{1}{\sqrt{\NS}} . 
  \label{eq:prod-uncertainty-propagation}
\end{align}
If there is no correlation in quantum system, then the readout features for both quantum moment readout and classical polynomial readout are the same $\langle \hat{\sigma}_{l_1}^z \hat{\sigma}_{l_2}^z \cdots \hat{\sigma}_{l_m}^z \rangle = \langle \hat{\sigma}_{l_1}^z \rangle \langle \hat{\sigma}_{l_2}^z \rangle \cdots \langle \hat{\sigma}_{l_m}^z \rangle$. However, even if the expectations under infinite sampling limit $\NS \rightarrow \infty$ are the same, the measurement noise under finite sampling are still different. For classical polynomial readout, the scaling of still follows the simple additivity relation of uncertainty propagation in Eq.\,(\ref{eq:prod-uncertainty-propagation}). But now the noise of $x_{l_1} (u) \cdots x_{l_m} (u)$ in quantum moment readout will be very strong, this is because $x_{l_1} (u) \cdots x_{l_m} (u)$ is now close to zero, thus
\begin{align}
    \left| \frac{\delta_{k}}{x_{k} (u)} \right| \approx \frac{1}{x_k (u)} \frac{1}{\sqrt{\NS}} = \frac{1}{x_{l_1} (u) \cdots x_{l_m} (u)} \frac{1}{\sqrt{\NS}} \propto 2^m \times \frac{1}{\sqrt{\NS}}.
\end{align}

\begin{figure}
    \centering
    \includegraphics[width=0.5\textwidth]{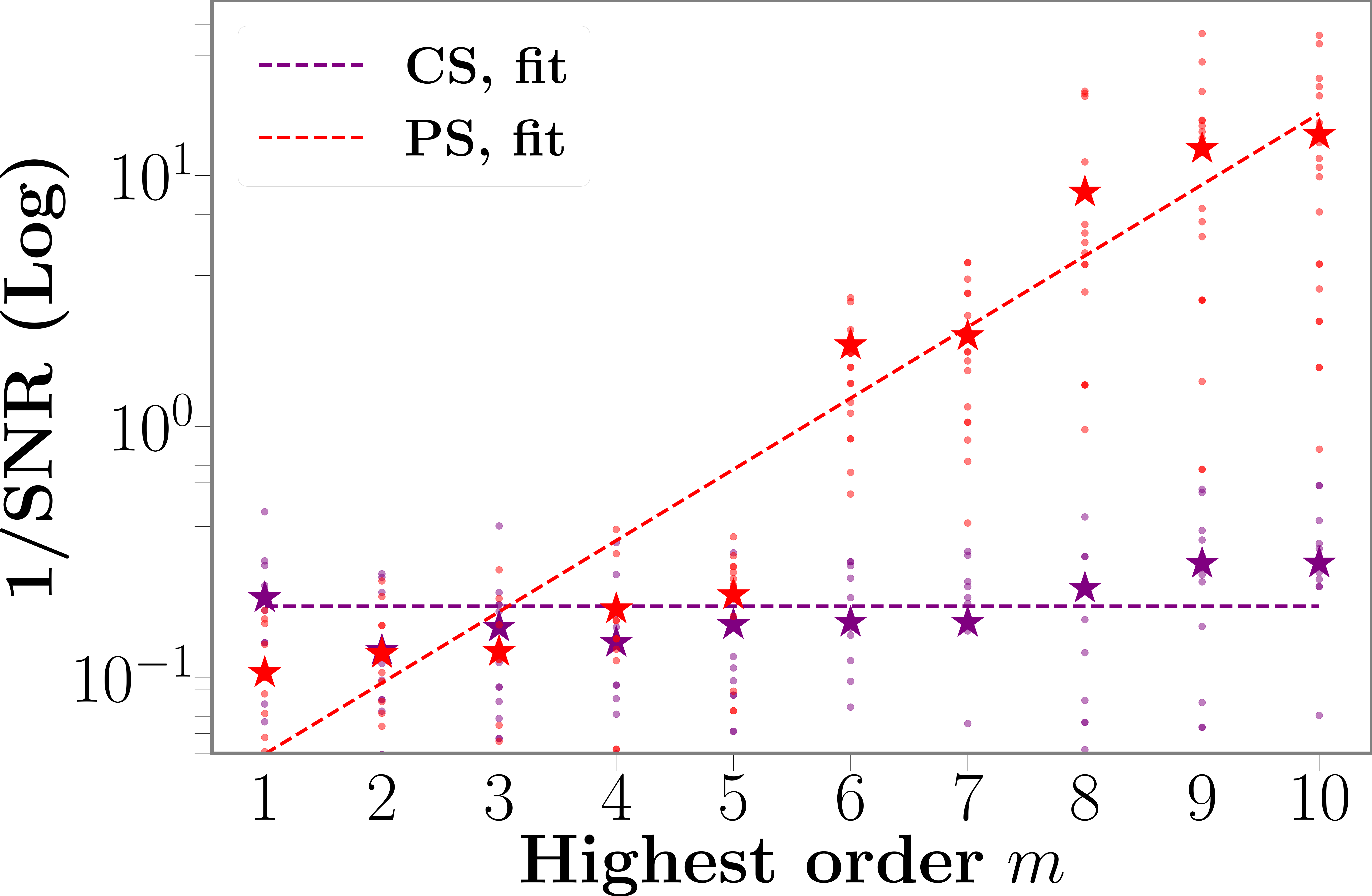}
    \caption{Noise-to-signal ratio of correlated system vs product system in a $10$-qubit quantum annealing system with shot number $\NS=1000$ by feeding $u=1/2$. The hyperparameters are chosen to be $(\bar{h}^x, h^x_{\mathrm{rms}}; \bar{h}^z, h^z_{1,\mathrm{rms}}) = (8, 2; 3, 2)$ in unit $1/t$. The purple and red colors correspond to coupling being switched on and off, respectively; and the coupling hyperparameter in CS is $J_{\mathrm{max}}=2/t$. For each $m$, the $N=30$ dots are relative error $x^{(r)}_{k}(u)/x_{k}(u)-1$ of 30 repetitions $r=1,2,\cdots,30$. The standard deviation of those relative errors (namely NSR) are also plotted. The correlated system NSR (purple stars) is well fitted by $O(1/\sqrt{\NS})$ (purple dashed line) while the product system NSR (red stars) scales exponentially as $O(2^m/\sqrt{\NS})$ (purple dashed line). We take $y$-axis being log-scale, and one may find in these regime correlated system 1/SNR grows exponentially faster than product system NSR (red stars) and hence product system readout scheme will be less powerful in sense of quantum sampling noise resistant. }
    \label{fig:EC_vs_PC}
\end{figure}


\end{widetext}

\end{document}